\documentclass{goose-article}

\hypersetup{pdfauthor={%
  T.W.J. de Geus%
}}
\header{%
  T.W.J.~de~Geus, R.H.J.~Peerlings, M.G.D.~Geers\\%
  International Journal of Solids and Structures, 2015, 67-68:326-339, %
  \doi{10.1016/j.ijsolstr.2015.04.035}%
}

\title{%
  Microstructural topology effects on the onset of ductile failure in multi-phase materials -- a systematic computational approach%
}
\author[1,2]{T.W.J.~de~Geus$^*$}
\author[1]{R.H.J.~Peerlings}
\author[1]{M.G.D.~Geers}
\affil[1]{%
  Department of Mechanical Engineering, Eindhoven University of Technology, Eindhoven, The Netherlands%
}
\affil[2]{%
  Materials innovation institute (M2i), Delft, The Netherlands%
}
\contact{%
  $^*$Corresponding author:
  \href{mailto:t.w.j.d.geus@tue.nl}{t.w.j.d.geus@tue.nl} %
  \hspace{1mm}--\hspace{1mm} %
  \href{mailto:tom@geus.me}{tom@geus.me} %
  \hspace{1mm}--\hspace{1mm} %
  \href{http://www.geus.me}{www.geus.me}%
}

\begin{document}

\maketitle

\begin{abstract}
Multi-phase materials are key for modern engineering applications. They are generally characterized by a high strength and ductility. Many of these materials fail by ductile fracture of the, generally softer, matrix phase. In this work we systematically study the influence of the arrangement of the phases by correlating the microstructure of a two-phase material to the onset of ductile failure. A single topological feature is identified in which critical levels of damage are consistently indicated. It consists of a small region of the matrix phase with particles of the hard phase on both sides in a direction that depends on the applied deformation. Due to this configuration, a large tensile hydrostatic stress and plastic strain is observed inside the matrix, indicating high damage. This topological feature has, to some extent, been recognized before for certain multi-phase materials. This study however provides insight in the mechanics involved, including the influence of the loading conditions and the arrangement of the phases in the material surrounding the feature. Furthermore, a parameter study is performed to explore the influence of volume fraction and hardness of the inclusion phase. For the same macroscopic hardening response, the ductility is predicted to increase if the volume fraction of the hard phase increases while at the same time its hardness decreases.
\end{abstract}


\section{Introduction}

\subsection{Background}

Multi-phase materials are frequently used in engineering applications because they generally provide a good compromise between a high strength and a high ductility. Examples are dual-phase steels, metal matrix composites, and fiber-reinforced polymers. For many of these multi-phase materials the macroscopic response as a function of the microstructure has been reasonably well characterized, both experimentally and numerically \citep{Choi2009, Davies1978, Sun2009, Deng2006, Heinrich2012, Brockenbrough1990, Povirk1995}. Existing models range from simple phenomenological models to complex multi-scale simulations. So far, it is only partially understood which microstructural mechanisms govern the failure of these materials \citep{Tasan2009, Choi2009a, Uthaisangsuk2008a, Williams2010, Babout2004}. To better understand and characterize the failure of multi-phase materials one needs to consider the response at the microstructural level; the level at which the inhomogeneity between phases is clearly distinguishable, and damage initiates.

\subsection{State of the art}

Experimental evidence exists that suggests that failure often occurs by ductile fracture of the, generally relatively soft, matrix phase. This is substantiated for example by fractography, in-situ microstructural observations using Scanning Electron Microscopy (SEM), or X-ray tomography \citep{Kadkhodapour2011, Avramovic-Cingara2009a, Avramovic-Cingara2009, Ahmad2000, Tasan2010, Williams2010}. Similar observations are made using models \citep{Asgari2009, Kadkhodapour2011, Sun2009a}. However, depending on the configuration and loading conditions considered, also brittle failure of the hard inclusion phase or decohesion of the interface between the two phases has been observed \citep{Ahmad2000, Avramovic-Cingara2009, Avramovic-Cingara2009a, Uthaisangsuk2008}.

To better understand the onset of failure in multi-phase materials, a wide variety of microstructural models are used. These models often use relatively simple material representation in which both phases are modeled using an elasto--plastic constitutive model and where failure is associated with large local permanent deformation \citep[e.g.][]{Choi2009, Choi2009a, Sun2009, Sun2009a, Kumar2006, Povirk1995}. Sometimes, more complicated models are used to quantify the mechanisms underlying ductile fracture. For example, Prahl and co-workers applied the Gurson--Tvergaard--Needleman (GTN) model \citep{Prahl2007, Uthaisangsuk2008}. Using this extension it was shown that for transformation and twinning induced plasticity, void nucleation is due to phase transformations or the presence of inclusions. The competition between these two mechanisms is largely influenced by the stress triaxiality.

The mechanism leading to final failure is however not easily predicted. Different failure modes, ranging from shear to split failure, are observed for different loading conditions \citep{Choi2009, Choi2009a, Sun2009}. Less apparent is the role of the microstructure on the onset of these mechanisms. So far, \citet{Choi2009a} established that lower levels of damage occur when the hard phase is distributed more homogeneously, since stress concentrations are relaxed \citep[see also][]{Huang2008, Segurado2003}. A more systematic approach was proposed by \citet{Kumar2006}. Their objective was to identify a microstructural feature in which high levels of damage occur, independent of topological variation outside this feature. In their preliminary conclusion they identify a damage ``hot-spot'' in which a high level of damage is observed in the soft phase with hard phase on both sides in the principal strain direction.

\subsection{Objective}

Except \citet{Kumar2006}, none of the described studies perform a systematic analysis of the influence of microstructural topology. In fact, many studies use realistic microstructures, often directly obtained from experiments, thus limiting the space for carrying out a systematic analysis. Such a systematic study is however of interest for multi-phase materials since the presence of at least two distinct phases raises the question how these phases should be arranged to optimally benefit from their distinct properties. Therefore, in this paper focuses on the influence of the microstructural topology on damage initiation in a ductile matrix, for a wide variety of multi-phase materials. To this end, a systematic analysis is performed in which a large number of randomly generated microstructures are compared. The microstructural model that is used is highly idealized in order to enable the identification of the influence of the geometrical arrangement of the phases independently from other influences -- which is not easily achieved experimentally or with realistic microstructures. By comparing a large ensemble of microstructures, the influence of the microstructural topology on the onset of ductile fracture is identified more clearly than by \citet{Kumar2006}, where a single microstructure was used as the basis of the analysis.

\subsection{Approach followed}

The microstructural model is two-dimensional (plane strain) and comprises square hard particles which are randomly distributed in a soft matrix phase consisting of equally sized elements (or grains). Both phases are modeled by isotropic elasto-plasticity. Given this highly idealized geometry and material model, only element averaged stresses and strains are considered. The onset of failure of the ductile matrix is predicted using a simple damage indicator which incorporates the effect of plastic strain and hydrostatic stress \citep[cf.][]{Freudenthal1950, McClintock1968, Rice1969}.

An advantage of the simple microstructures and damage model used, is that the resulting analyses are computationally inexpensive enabling the study of a large number of randomly generated microstructures. Unlike studies which are aimed at understanding, or predicting, the macroscopic material response before failure, the focus is here on the `worst case scenarios' -- i.e. geometrical features in the microstructure which give rise to high values of damage. The underlying reasoning is that in real materials such features will always be present somewhere in the material's microstructure and they are therefore preferential locations at which failure will initiate.

Through a novel statistical comparison an ensemble of microstructures is considered at once. The outcome of this analysis suggests that a single topological feature is critical in terms of the initiation of ductile fracture. It consists of a small region of the matrix phase, flanked by hard phase on both sides in the direction of the maximum principal strain, and by matrix material in the perpendicular direction. Our simulations consistently indicate the highest levels of damage in the soft matrix region at the center of this feature. By studying the underlying mechanics, we show that the orientation of the feature with respect to the direction of straining is essential. Given this critical feature and orientation, the level of damage is determined by the topology of the microstructure in the area directly around it. The influence of the volume fraction and hardness of the hard inclusion phase are also studied.

\subsection{Outline}

The outline of this paper is as follows. The microstructural model that is employed in our study is discussed in Section~\ref{sec:model}. Simulation results of the model are presented in Section~\ref{sec:ref}, resulting in the identification of a critical topological feature. The mechanics involved in this feature are also discussed in this section. In Section~\ref{sec:param}, a parameter study is performed to identify the influence of volume fraction and hardness of the inclusion phase on the initiation of failure. The paper closes with a discussion in Section~\ref{sec:discussion} and conclusions in Section~\ref{sec:conclusion}.

\section{Modeling}
\label{sec:model}

\subsection{Microstructure}

\begin{figure}[htp]
\centering
\includegraphics[width=130mm]{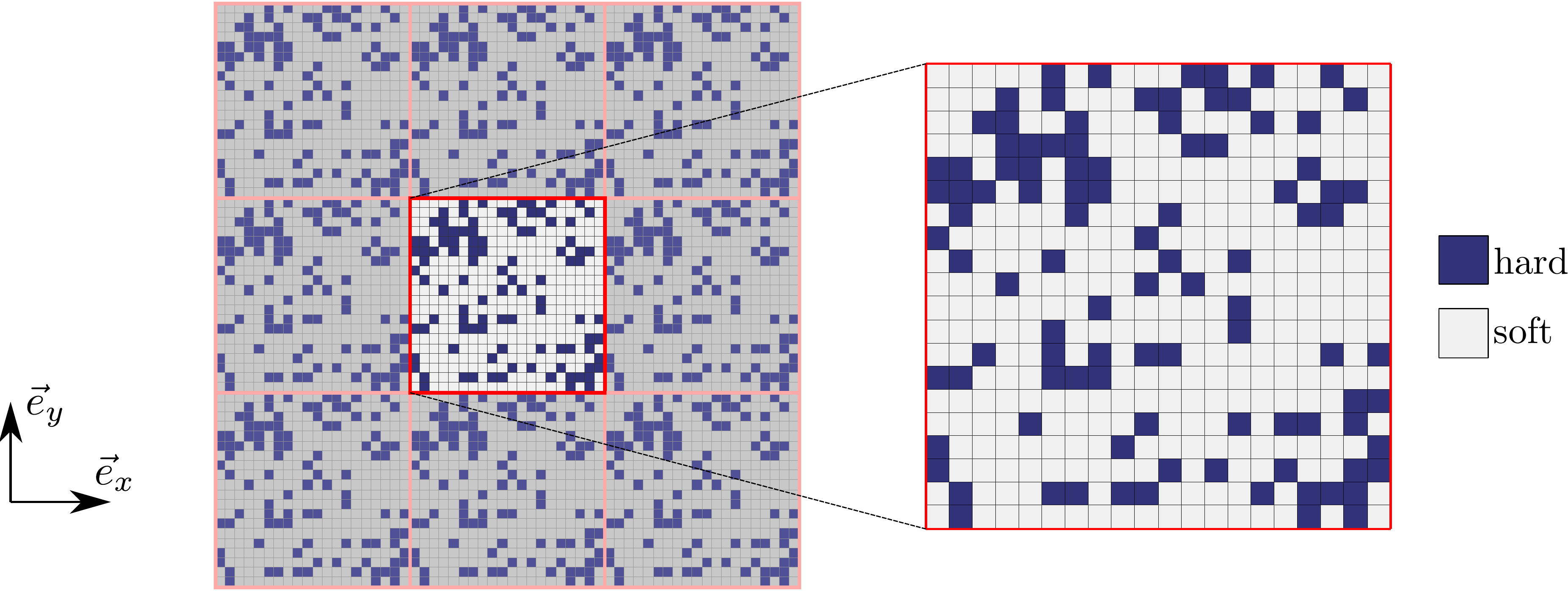}
\caption{Periodic cell of the random two-phase material.}
\label{fig:rve}
\end{figure}

Multi-phase materials consist of several phases, each with their own properties, generally in a highly disordered microstructure with irregularly shaped three-dimensional particles and grains of varying sizes. Here, the focus is on a two-phase material in which a harder inclusion phase is embedded in a softer matrix phase. Since we are mainly interested in the influence of the spatial arrangement of the two phases, a highly idealized two-dimensional microstructure is considered which consists of square, equi-sized elements (see Figure~\ref{fig:rve}). The interpretation of these elements depends on the class of materials considered, e.g.\ grains in multi-phase steels, or continuum volume elements in fiber reinforced polymers.

The microstructure is assumed to be periodic, whereby the periodicity is much larger than the individual elements. As a result, the entire microstructure may be modeled by a periodic cell to which periodic boundary conditions are applied; see Figure~\ref{fig:rve}. The microstructure shown in this figure is just a single realization of the large reference set of realizations used throughout this paper, for the statistical analysis. It consists of $20 \times 20$ microstructural elements, in which hard particles are randomly distributed according to a constant volume fraction $\varphi^\mathrm{hard} = 0.25$. Different topologies are obtained by analyzing $n_\mathrm{cell} = 400$ randomly generated microstructures (or unit cells).

\subsection{Constitutive model}

We assume that both phases obey an isotropic elasto-plastic constitutive behavior, but with different initial yield stresses and hardening laws. As the local deformations are significant we assume finite deformations. The elastic response is characterized by the Young's modulus $E$ and Poisson's ratio $\nu$. For the plastic part, standard Von Mises plasticity is used. The plastic hardening is described by a power-law, i.e.\ the yield stress is given by
\begin{equation}
  \label{eq:model:yield}
  \sigma_\mathrm{y} = \sigma_\mathrm{y0} + H \varepsilon_\mathrm{p}^n
\end{equation}
where $\varepsilon_\mathrm{p}$ is the effective plastic strain. The parameters $\sigma_\mathrm{y0}$, $H$, and $n$ are the initial yield stress, hardening modulus and hardening exponent, respectively. They are defined for the two phases as follows
\begin{equation}
  \label{eq:model:param}
  \frac{\sigma_\mathrm{y0}^\mathrm{hard}}{\sigma_\mathrm{y0}^\mathrm{soft}} = 3,
  \quad
  \frac{H^\mathrm{hard}}{\sigma_\mathrm{y0}^\mathrm{soft}} = 4,
  \quad
  n^\mathrm{hard} = 1;
  \quad
  \frac{H^\mathrm{soft}}{\sigma_\mathrm{y0}^\mathrm{soft}} = 2,
  \quad
  n^\mathrm{soft} = 0.2
\end{equation}
These parameters have been taken from one of the material classes for which our analysis is relevant: dual-phase steels \citep[see e.g.][]{Sun2009}. However, the level of property contrast resulting from these parameter values is relevant for a wider class of materials, e.g.\ for metal matrix composites.

\subsection{Applied deformation}

The (periodic) cell is loaded in pure shear, in combination with a plane strain condition for the out of plane direction. The macroscopic -- unit cell averaged -- deformation gradient tensor therefore reads
\begin{equation}
  \label{eq:model:def}
  \bar{\bm{F}} = \lambda \, \vec{e}_x \vec{e}_x +
  \frac{1}{\lambda} \vec{e}_y \vec{e}_y + \vec{e}_z \vec{e}_z
\end{equation}
where $\lambda \geq 0$ is the stretch factor in the $\vec{e}_x$ direction. The deformation is applied monotonically, until a final stretch factor of $\lambda = 1.2$. We characterize the macroscopic deformation by the logarithmic strain tensor
\begin{equation}
  \bar{\bm{\varepsilon}} =
  \ln \lambda \, \big( \vec{e}_x \vec{e}_x - \vec{e}_y \vec{e}_y \big)
\end{equation}
and the effective (logarithmic) strain
\begin{equation}
  \bar{\varepsilon} = \frac{2}{\sqrt{3}} \ln \lambda
\end{equation}

\subsection{Failure criterion}

We next focus on the onset of ductile fracture of the soft and ductile matrix material. Failure of the hard and brittle inclusion phase is not considered. The high degree of idealization of the microstructures used implies that local peak stresses and strains, e.g.\ at element corners, should not be considered as physically correct fluctuations. Therefore, only element averaged stresses and strains are considered, which are less sensitive to these local fluctuations and to the precise geometry of the elements. The criterion for the onset of failure is also formulated in terms of these averaged stresses and strains.

It is well known that ductile failure occurs through a mechanism of void nucleation, growth, and finally coalescence into a macroscopic crack. Many studies have been devoted to understanding and describing these mechanisms. Different models incorporating the influence and evolution of voids in the constitutive model have been proposed \citep[e.g.][]{Argon1975, Beremin1981, McClintock1968, Rice1969, Gurson1977}. A common feature of these models is that for voids to nucleate and grow, a hydrostatic tensile stress state (i.e.\ a positive mean stress, $\sigma_\mathrm{m} > 0$) is required. In contrast, when $\sigma_\mathrm{m} < 0$ any existing void would be closed. Furthermore, void growth is accompanied by a significant amount of permanent deformation of the matrix material.

A simple criterion is used for the onset of ductile failure in terms of a qualitative damage indicator $D$, which depends on the effective plastic strain and the hydrostatic stress as follows
\begin{equation}
  \label{eq:model:damage}
  D = \varepsilon_\mathrm{p} \, \lfloor \sigma_\mathrm{m} \rfloor
\end{equation}
where the brackets $\lfloor\ldots\rfloor$ denote that only non-negative values are taken into account, i.e.\
\begin{equation}
  \lfloor x \rfloor \, = \tfrac{1}{2} \big( \, x + |x| \, \big)
\end{equation}
and $\varepsilon_\mathrm{p}$ and $\sigma_\mathrm{m}$ are the element-averaged effective plastic strain and hydrostatic stress respectively. This damage variable may be regarded as a simplified version of a void growth model. It is used solely as an indicator to identify elements subjected to stress and plastic strain states that are likely to result in ductile failure. Computed damage values have no effect on the constitutive response of the material, i.e.\ the damage variable is not coupled to the elasto-plasticity model.

\subsection{Implementation}

The response of the unit cells (with a randomly distributed microstructure) is calculated using the finite element method. Each microstructural element is discretized by $5 \times 5$ quadratic finite elements, so that the complete unit cell consists of $100 \times 100$ finite elements (see Figure~\ref{fig:rve}). Reduced integration is applied so that each finite element has four integration points. A mesh refinement study -- not included in this paper -- shows that this discretization is sufficiently fine to yield converged element-averaged quantities. Tyings are used between opposite boundaries of the cell to implement the periodic boundary conditions. achieved imposed deformation is applied in $2000$ increments and convergence is tested using a relative displacement tolerance of~$10^{-3}$.

\subsection{Notation}

Throughout this paper, we distinguish between volume and ensemble averages; the averaging on microstructural elements (grains) as discussed above is always applied and therefore not explicitly denoted. The unit cell average of a quantity $a$ is denoted by $\bar{a}$. The average on the ensemble -- i.e.\ the 400 random unit cells -- for a particular element on the other hand is denoted $\langle a \rangle$. $\langle \bar{a} \rangle$ indicates the combined operation, i.e.\ the ensemble average of the volume average.

\section{Simulation results}
\label{sec:ref}

\subsection{Macroscopic response}

The average macroscopic response of the 400 different, randomly generated, microstructures is illustrated first. The macroscopic Von Mises equivalent stress $\bar{\sigma}_\mathrm{eq}$ is calculated from the macroscopic Cauchy stress tensor determined by volume averaging of the local stress tensors over the entire cell. In Figure~\ref{fig:ref:macros} the ensemble average $\langle \bar{\sigma}_\mathrm{eq} \rangle$ is plotted as a function of the macroscopic applied equivalent strain $\bar{\varepsilon}$. The data has been normalized by the initial yield stress of the soft phase, $\sigma_\mathrm{y0}^\mathrm{soft}$, and the initial yield strain $\varepsilon_0^\mathrm{soft}$ of the soft phase, where the latter is defined as $\varepsilon_0^\mathrm{soft} = \sigma_\mathrm{y0}^\mathrm{soft} / E$.  In the diagram, the homogeneous response of the hard and soft phase is included as an upper and lower bound respectively.

\newcommand{\markeri}{}
\newcommand{\markerj}{}
\newcommand{\markerk}{}
\renewcommand{\markeri}[0]{%
  \protect\includegraphics[height=.5em]{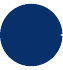}%
}%
\renewcommand{\markerj}[0]{%
  \protect\includegraphics[height=.5em]{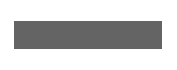}%
}%
\renewcommand{\markerk}[0]{%
  \protect\includegraphics[height=.5em]{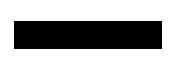}%
}%
\begin{figure}[htp]
\centering
\includegraphics[width=70mm]{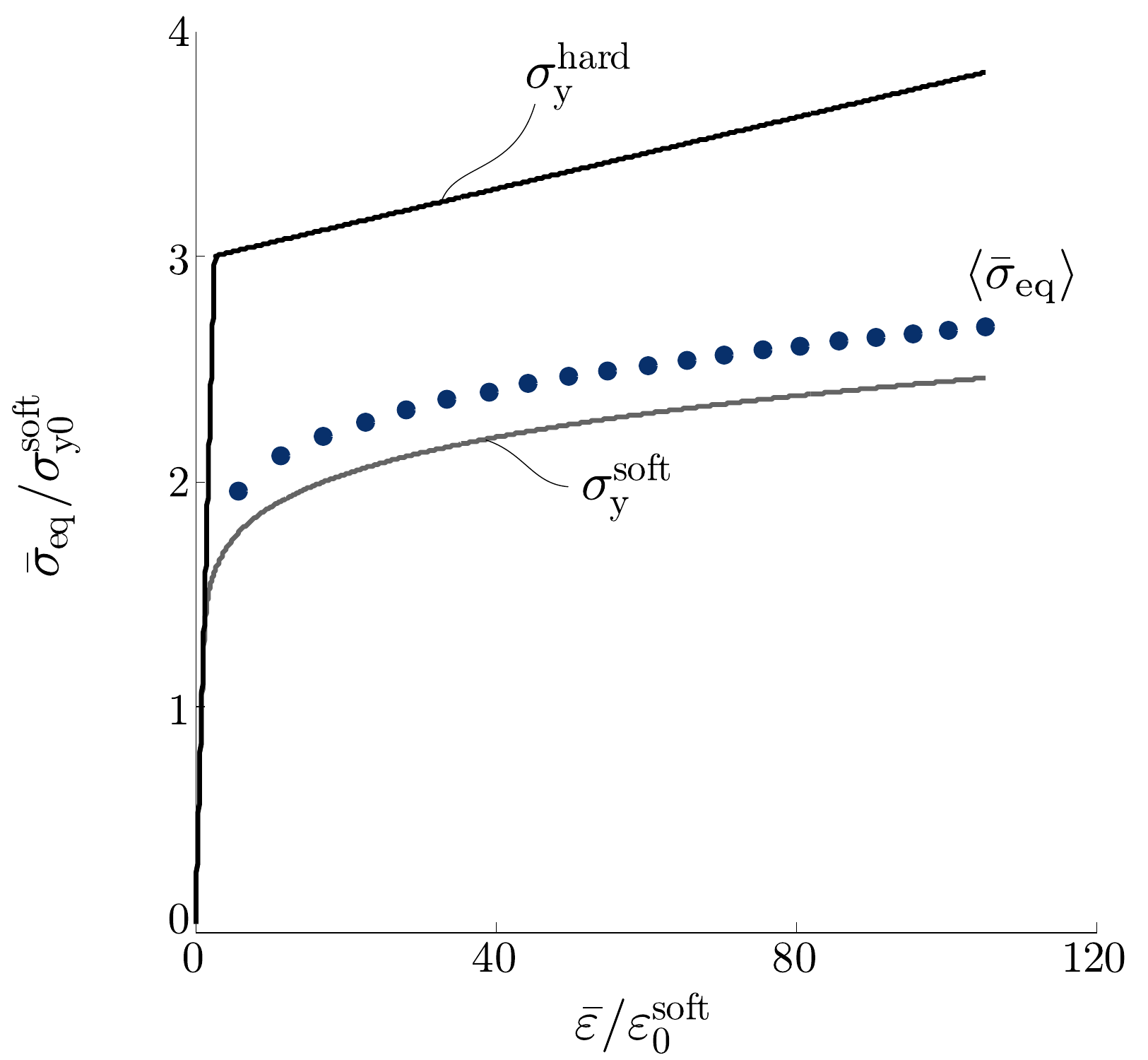}
\caption{Ensemble averaged Macroscopic Von Mises equivalent stress $\langle \bar{\sigma}_\mathrm{eq} \rangle$ (\markeri) versus the macroscopic applied equivalent strain $\bar{\varepsilon}$, respectively normalized by the initial yield stress and strain, $\sigma_\mathrm{y0}^\mathrm{soft}$ and $\varepsilon_0^\mathrm{soft}$, of the soft matrix. The homogeneous stress of the soft phase (\markerj) and the hard phase (\markerk) act as a lower and upper bound respectively.}
\label{fig:ref:macros}
\end{figure}

\subsection{Scatter of the overall response and damage}

\begin{figure}[htp]
\centering
\begin{minipage}[b]{75mm}
  \includegraphics[width=1.\textwidth]{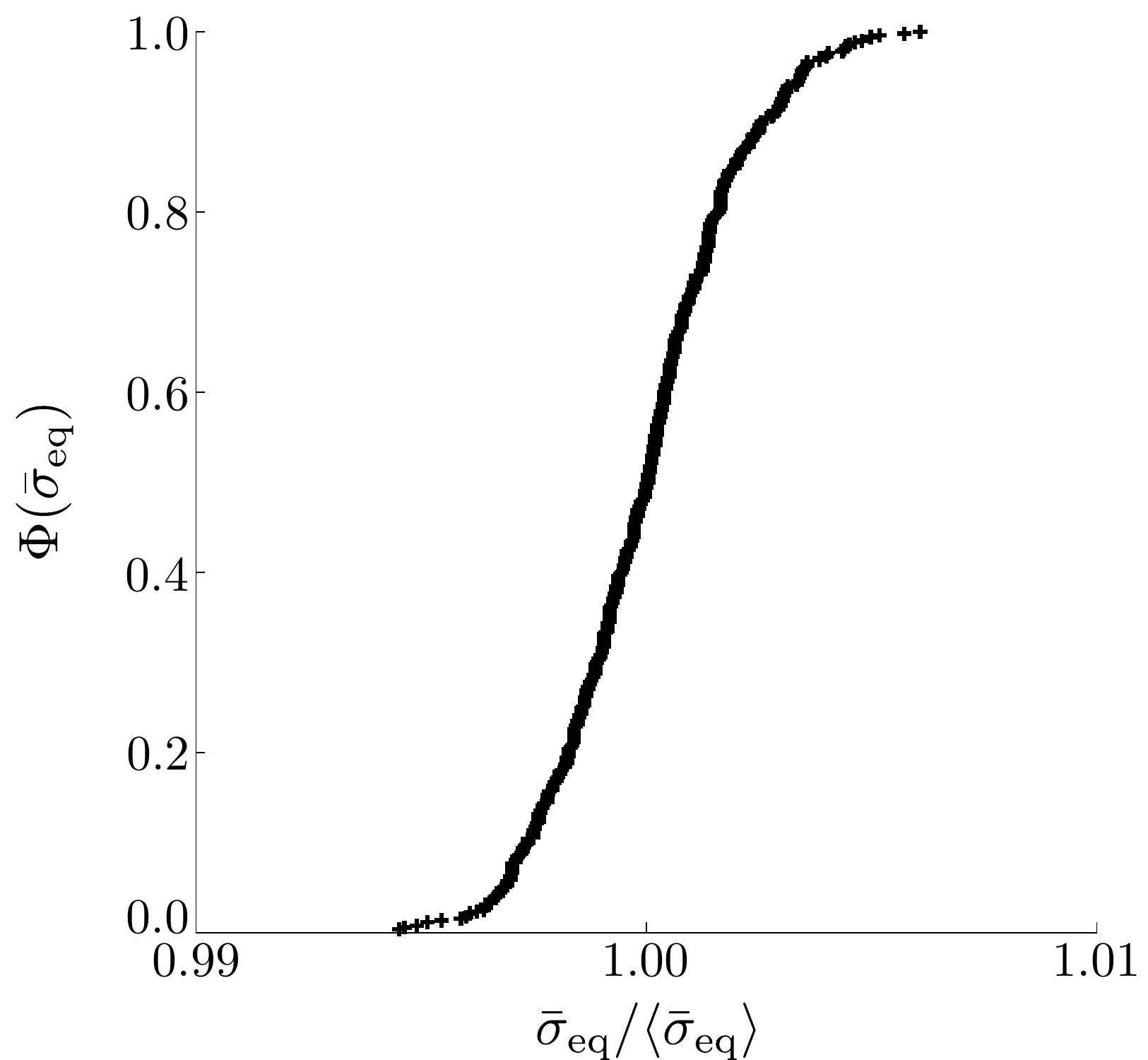}
  \\
  \hspace*{12.9mm}%
  \footnotesize (a) macroscopic stress
\end{minipage}
\hspace{5mm}
\begin{minipage}[b]{75mm}
  \includegraphics[width=1.\textwidth]{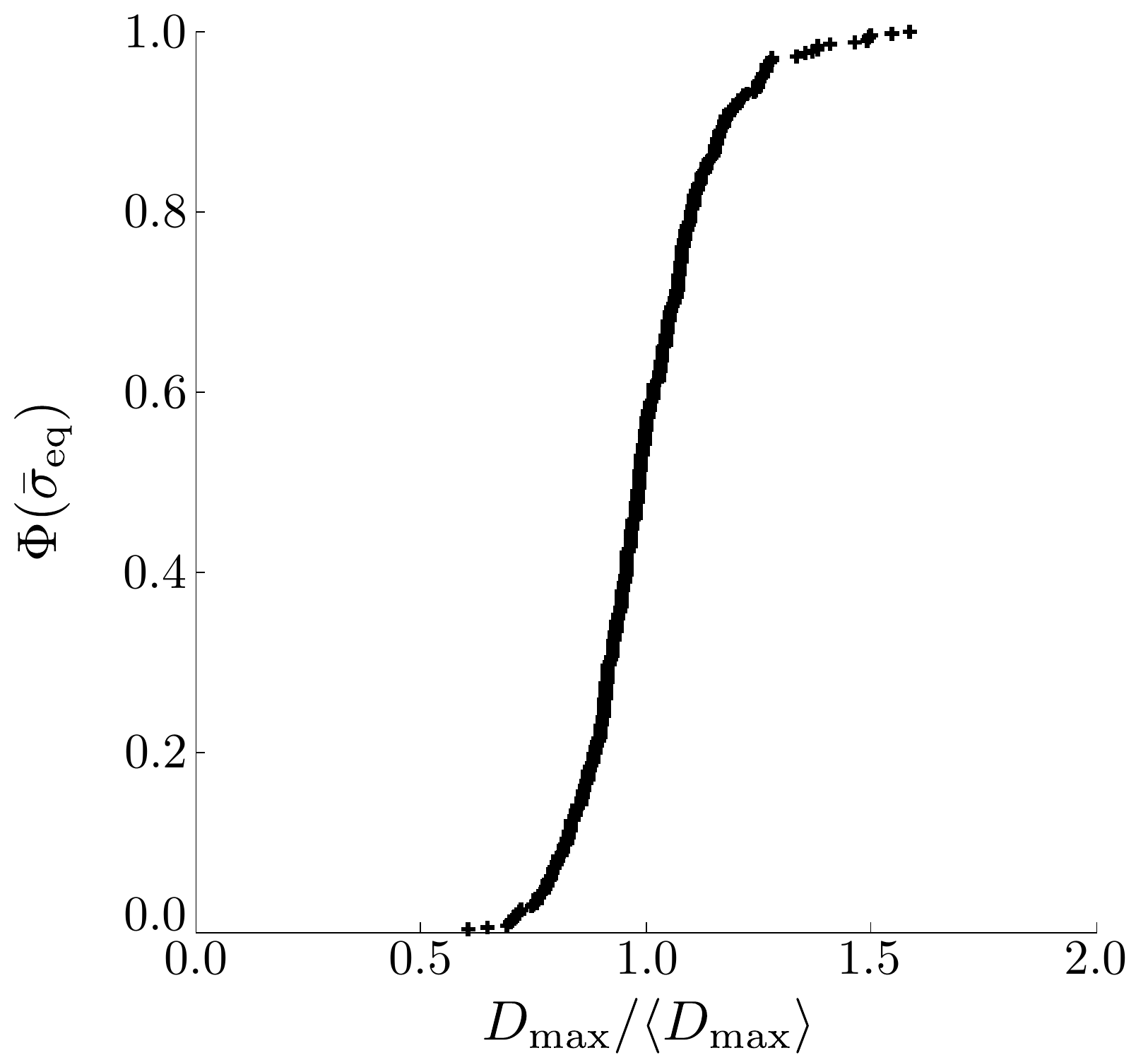}
  \\
  \hspace*{12.9mm}%
  \footnotesize (b) damage
\end{minipage}
\caption{Cumulative distribution of (a) the macroscopic Von Mises equivalent stress $\bar{\sigma}_\mathrm{eq}$ and (b) the maximum damage indicator $D_\mathrm{max}$ per cell, normalized by their respective ensemble averages $\langle \bar{\sigma}_\mathrm{eq} \rangle$ and $\langle D_\mathrm{max} \rangle$; both diagrams are for an applied deformation of $\lambda = 1.2$. Notice the large difference in horizontal scales.}
\label{fig:ref:scatter}
\end{figure}

The scatter in the responses of the individual cells around the average shown in Figure~\ref{fig:ref:macros} is small. This is demonstrated through the cumulative distribution of the macroscopic, or cell averaged, equivalent stress at the final deformation. In Figure~\ref{fig:ref:scatter}(a), the cumulative probability density $\Phi$ of $\bar{\sigma}_\mathrm{eq}$ is plotted, whereby the latter is normalized by the ensemble average, $\langle \bar{\sigma}_\mathrm{eq} \rangle$, corresponding to the final data point in Figure~\ref{fig:ref:macros}. Obviously, the influence of the microstructural topology on the macroscopic response is negligible, since the scatter is within $1\%$ of the average $\langle \bar{\sigma}_\mathrm{eq} \rangle$. This implies that the macroscopic response of each individual cell is representative for the entire ensemble. Such a periodic cell is commonly denoted as a representative volume element (RVE). A similar conclusion was drawn for dual-phase steels \citep[e.g.][]{Choi2009, Sun2009}, for metal matrix composites \citep[e.g.][]{Deng2006}, and for fiber-reinforced polymers \citep[e.g.][]{Huang2008, Heinrich2012}, provided that the unit cell contains sufficient elements.

Similar to the macroscopic stress, in Figure~\ref{fig:ref:scatter}(b) a cumulative distribution is given for the maximum damage reached in each of the cells at the final deformation, $D_\mathrm{max}$. In this case a large scatter is observed, with an amplitude that is of the same order as the average $\langle D_\mathrm{max} \rangle$ (notice the different horizontal scales used in Figure~\ref{fig:ref:scatter}). This highlights the influence of the microstructural topology on damage. In contrast to the macroscopic equivalent stress, the cells are not representative in terms of damage. For this reason we consider many cells with different distributions of the phases.

\subsection{Individual responses}
\label{sec:ref:individual}

\begin{figure}[htp]
\centering
\includegraphics[height=130mm]{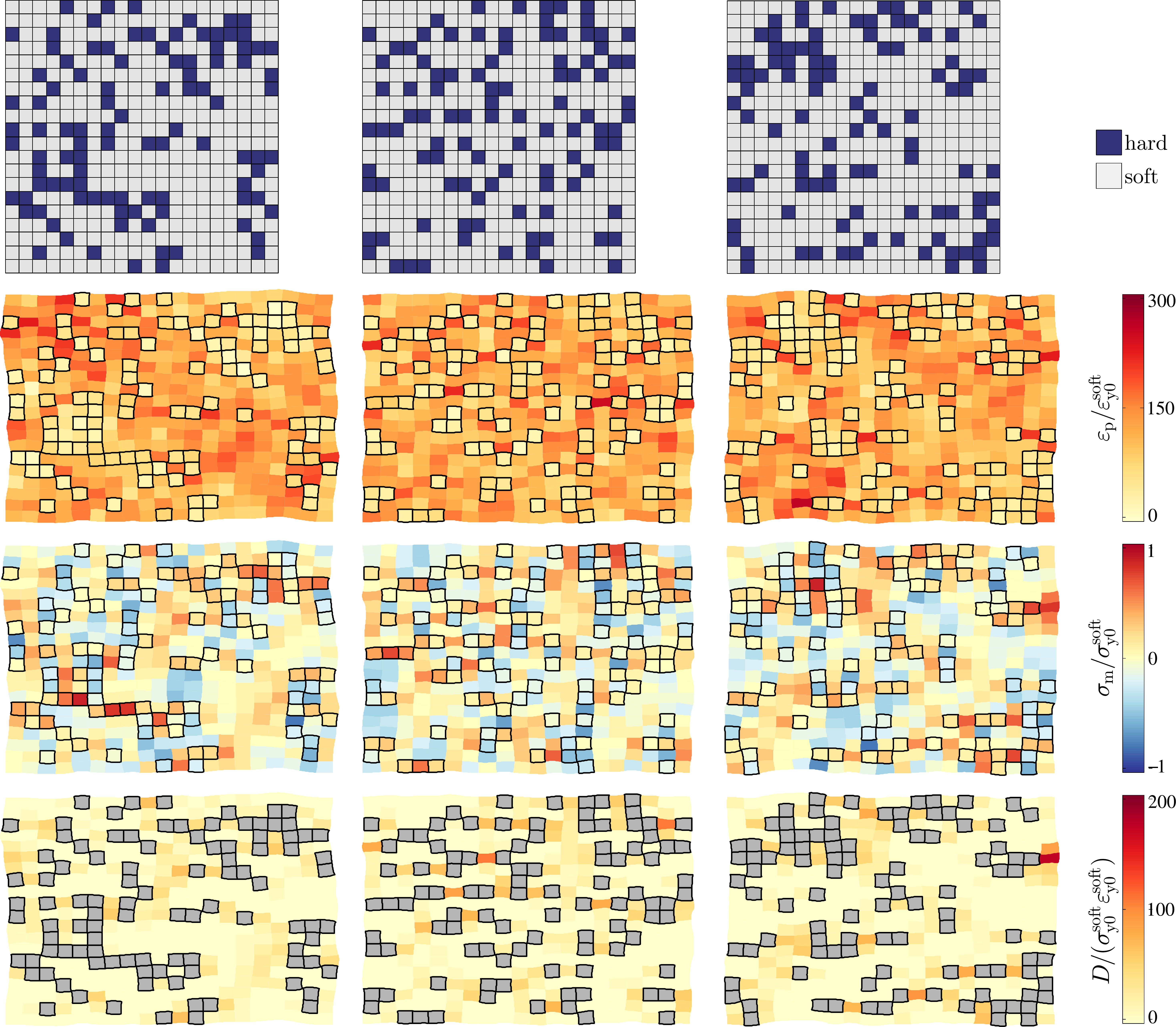}
\\ \vspace{0.3eM}
\footnotesize%
\begin{minipage}{38mm}
  (a) lowest damage%
\end{minipage}%
\hspace{7.5mm}%
\begin{minipage}{38mm}
  (b) intermediate damage%
\end{minipage}%
\hspace{7.5mm}%
\begin{minipage}{38mm}
  (c) highest damage%
\end{minipage}%
\hspace*{17mm}%
\caption{Simulation results of three randomly generated cells. From left to right, the maximum damage in the cell is the (a) lowest, (b) intermediate, and (c) highest among all cells. From top to bottom: the microstructure, equivalent plastic strain $\varepsilon_\mathrm{p}$, hydrostatic stress $\sigma_\mathrm{m}$, and damage $D$ at $\lambda = 1.2$. The hard phase (i.e.\ inclusions) is highlighted by a black outline.}
\label{fig:ref:response}
\end{figure}

To assess the local configurations triggering damage, three different cells on both extremes and at the center of the distribution in Figure~\ref{fig:ref:scatter}(b) (i.e.\ $D_\mathrm{max} / \langle D_\mathrm{max} \rangle \approx 0.6$, $1.0$, and $1.6$) are analyzed. These microstructures and their response are shown in Figure~\ref{fig:ref:response}, in which the maximum damage increases from left to right. From top to bottom the figure shows -- for each of these cells -- the microstructure and the distribution of equivalent plastic strain $\varepsilon_\mathrm{p}$, hydrostatic stress $\sigma_\mathrm{m}$, and ductile damage indicator $D$ (only defined in the ductile soft phase).

In the plastic strain distributions shown in Figure~\ref{fig:ref:response}, shear bands develop at angles of approximately $45$ degrees with respect to the loading axis. These bands develop predominantly in the soft phase, as its plastic resistance is lowest. High stress levels are observed in regions where soft and hard phase are adjacent to each other.

The cell in Figure~\ref{fig:ref:response}(a) is characterized by a region with a large amount of connected soft matrix material at the bottom right. In this region a significant portion of the plastic strain is taken up. However, due to the absence of hard phase particles in this region, the hydrostatic stress remains low and hence little damage is predicted. In the top left part of the unit cell small regions of matrix material are connected to large islands of the inclusion phase. The resulting strain heterogeneity causes a large amount of local plastic deformation. However, in this case the hydrostatic stress is predominantly compressive, resulting in a low level of damage.

In the cell of Figure~\ref{fig:ref:response}(b) the hard phase is distributed more homogeneously. This causes the plastic deformation to be distributed over the entire cell. As a result, the strain heterogeneity is lower and therefore also the hydrostatic stress is distributed more uniformly. Similarly to the cell in Figure~\ref{fig:ref:response}(a) relatively high (positive) stresses are located only in the hard phase and relatively large plastic deformations occur only in the soft phase. Both phases are able to accommodate this without indicating significant levels of damage.

In the cell of Figure~\ref{fig:ref:response}(c), a large amount of plastic deformation again develops inside the soft phase. Locally, high hydrostatic stresses develop due to the vicinity of hard particles. In particular where the soft phase is adjacent (left and right) to the hard inclusion phase, damage reaches a high level; note that, due to the periodic boundary conditions, the volume element is repeated in all directions.

\subsection{Statistical analysis of the topology influence}

To more rigorously identify the influence of the topology on damage, the likelihood of finding a particular phase at a certain position relative to the fracture initiation sites is quantified. Since only two phases are considered, it is sufficient to quantify the probability of finding the hard phase. If at a relative position the probability of hard phase is higher than the volume fraction, $\varphi^\mathrm{hard}$, having hard phase at this relative position promotes fracture initiation in the reference element. Likewise, if the probability is lower than $\varphi^\mathrm{hard}$, having soft phase at that relative position promotes damage.

More precisely, the damage weighted average distribution of the phases around fracture initiation is computed. This is shown first at the level of a single unit cell and subsequently we average also across all unit cells in the ensemble. The distribution of the hard and soft phase in the cell is described using a phase indicator, defined as follows:
\begin{equation}
  \mathcal{I}(i,j) =
  \begin{cases}
    1 &\quad \mathrm{for}\, (i,j) \in \mathrm{hard}
    \\
    0 &\quad \mathrm{for}\, (i,j) \in \mathrm{soft}
  \end{cases}
\end{equation}
where $(i,j)$ denotes the position in the cell, which for the structured cell can be thought of as the row and column index of the grid of elements. Likewise, the damage indicator, $D(i,j)$, is known in every location, resulting from the local mechanical response. The average phase at a certain position $(\Delta i, \Delta j)$ relative to the fracture initiation sites can now be obtained by calculating the average phase indicator weighted with the damage indicator, as follows:
\begin{equation} \label{eq:average_distro_D}
  \mathcal{I}_D ( \Delta i , \Delta j ) = \frac{\sum_{i,j} D(i,j) \; \mathcal{I} ( i + \Delta i , j + \Delta j ) }{\sum_{i,j} D(i,j) }
\end{equation}
where $(i,j)$ loops over all elements in the unit cell, thereby taking the periodicity of the unit cell into account. By using the damage indicator of the reference element as a weight factor, the surroundings of a highly damaged element contribute much more to $\mathcal{I}_D$ than those of a less damaged element. While averaging, a pattern may thus emerge which characterizes the typical surroundings of highly damaged elements. Note that without this weight factor, one would expect to recover a uniform probability equal to the volume fraction. To be less sensitive to statistical fluctuations, the ensemble average is taken, i.e.\ the result is averaged for all unit cells, resulting in the ensemble weighted average $\langle \mathcal{I}_D \rangle$ as a function of the relative position $(\Delta i, \Delta j)$ with respect to the damage site.

\begin{figure}[htp]
\centering
\includegraphics[height=55mm]{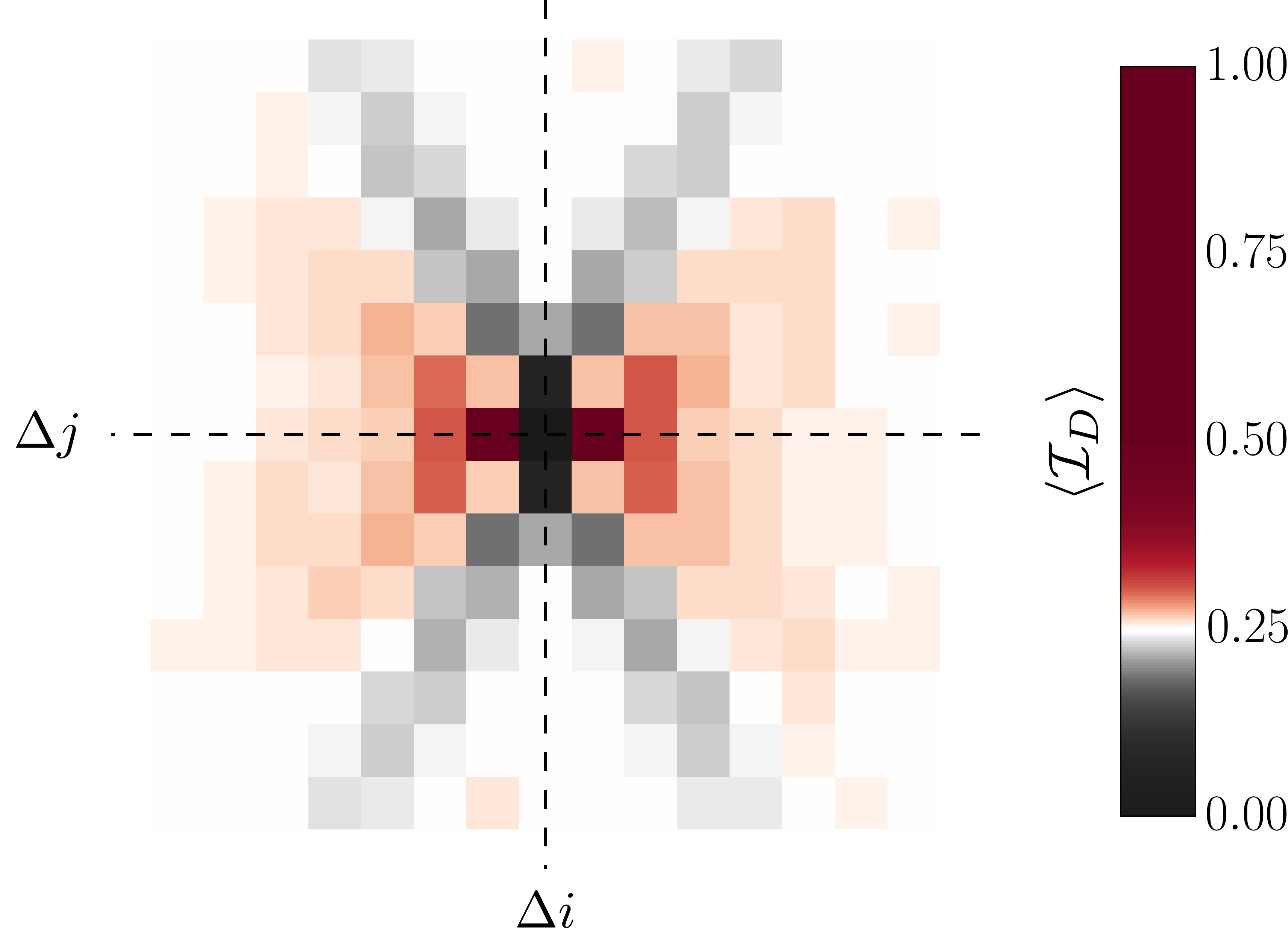}
\caption{The ensemble averaged indicator function weighed by the damage, $\langle \mathcal{I}_D \rangle$. As denoted by the axes, the origin is arbitrarily chosen in the center. The neutral color coincides with the volume fraction of the hard phase, $\varphi^\mathrm{hard} = 0.25$; red implies an elevated probability of finding hard phase whereas gray implies soft phase -- for $0.5 \leq \langle \mathcal{I}_D \rangle \leq 1.0$ a single color is used.}
\label{fig:ref:prob}
\end{figure}

The result is shown in Figure~\ref{fig:ref:prob}; note that the size of the region of interest shown in the figure is smaller than the unit cell. The color scale has been chosen such that the obtained phase patterns are highlighted. Following the theory above,
\begin{equation}
  \langle \mathcal{I}_D \rangle (\Delta i, \Delta j) \; = \; \varphi^\mathrm{hard}
\end{equation}
corresponds to no preferential phase, taken as the neutral -- white -- color in Figure~\ref{fig:ref:prob}.
\begin{equation}
  \varphi^\mathrm{hard} \; < \; \langle \mathcal{I}_D \rangle (\Delta i, \Delta j) \; \leq \; 1
\end{equation}
shown red in Figure~\ref{fig:ref:prob}, corresponds to an elevated probability of the hard phase at a position $(\Delta i, \Delta j)$ relative to sites with high damage; and
\begin{equation}
  0 \; \leq \; \langle \mathcal{I}_D \rangle (\Delta i, \Delta j) \; < \; \varphi^\mathrm{hard}
\end{equation}
shown gray in Figure~\ref{fig:ref:prob}, to an elevated probability of the soft phase at that relative position. It is remarked that the color varies in the range $(0,0.5)$ and is constant for $(0.5,1)$ in order to avoid an optical bias.

The result, in Figure~\ref{fig:ref:prob}, shows that in the immediate surrounding area the (centered) soft element is flanked to the left and right by hard elements, and by soft elements to the top and bottom -- see Figure~\ref{fig:crit}(b) for a sketch of this pattern. The orientation of this critical feature coincides with the direction of applied principal strain. Furthermore, an elevated probability of hard phase on both sides of this feature is observed aligned with the tensile deformation. An elevated probability of matrix phase is observed in $\pm 65$ degree angles\footnote{Figure~\ref{fig:ref:prob} is in the (undeformed) reference configuration, using the push-forward to the current configuration the matrix phase bands are approximately under $\pm 55$ degree angles.} through the critical feature. Far away from the central feature $\langle \mathcal{I}_D \rangle$ decays to the volume fraction $\varphi^\mathrm{hard}$, which is indicated in white in the figure. In other words, there is no preferential topology in these areas.

Note that \eqref{eq:average_distro_D} is constructed by linearly weighing the distribution with the damage indicator. The results do not significantly change when a different weighing is applied. In particular, a weighing has been applied in which for each cell only the maximum damage is taken into account, and the remaining elements are assigned zero weight. This leads to somewhat more scatter but essentially the same pattern as shown in Figure~\ref{fig:ref:prob}.

From the results above, including the simulation results discussed in Section~\ref{sec:ref}, it is clear that the highest damage occurs in the critical topological feature sketched in Figure~\ref{fig:crit}(b). The importance of such a feature has been observed previously in the literature \citep{Kadkhodapour2011, Kadkhodapour2011a, Williams2010}. It may also be regarded as a generalization of a small region of soft matrix interrupting a band of hard phase, which is known to be a critical location in dual-phase steel \citep{Tasan2010}. The exact orientation was however not identified previously.

The identified feature only leads to a high damage indicator when aligned unfavorably to the applied deformation. Indeed, when the feature is rotated $90$ degrees (i.e.\ when the hard particles are located above and below, instead of left of right, with respect to the central matrix region of matrix) damage is always zero, see e.g.\ the examples in Figure~\ref{fig:ref:response}. Furthermore, the damage level is strongly influenced by the surrounding topology in the vicinity of the feature. The latter explains why the maximum damage varies significantly among the cells of Figure~\ref{fig:ref:response}(a--c) even though each of them contains a number of critical features. Both of these influences -- i.e.\ the orientation of the feature, as well as of the topology in the vicinity of the feature -- are addressed in more detail in the next subsection.

\subsection{Mechanics of the critical feature}
\label{sec:crit:mech}

\begin{figure}[htbp]
\centering
\includegraphics[width=120mm]{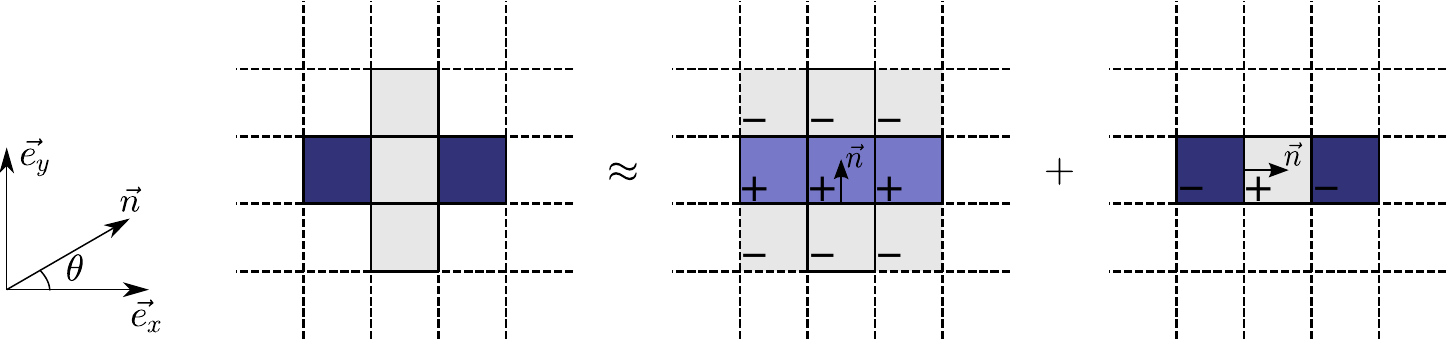}
\\
\begin{minipage}[b]{120mm}
  \footnotesize
  \hspace*{0.4mm}%
  (a)
  \hspace*{14.5mm}%
  (b)
  \hspace*{30.5mm}%
  (c)
  \hspace*{31.5mm}%
  (d)
\end{minipage}
\caption{Approximation of the critical feature, in (b), using a superposition of laminates, in (c--d). The resulting sign of the hydrostatic stress as a consequence of the orientation of the interface is indicated using $+/-$ symbols. The orientation of the interface is defined using the interface normal, $\vec{n}$, whose orientation with respect to the Cartesian basis is defined in (a).}
\label{fig:crit}
\end{figure}

In the critical feature, a small region of matrix has to accommodate most of the local plastic deformation, as the neighboring inclusion phase particles are (much) harder. Previously, a tensile hydrostatic stress has been observed in the critical feature of Figure~\ref{fig:crit}(b). This stress state is caused by the phase boundary separating the hard and soft phase. This can be shown by evaluating the stress state in a two-phase laminate of which the phases are perfectly bonded. The interface between the phases is characterized by a normal vector $\vec{n} = \cos\theta \, \vec{e}_x + \sin\theta \, \vec{e}_y$, where $\theta$ equals either $0$ or $90$ degrees. Consider such a laminate, in which we neglect elastic deformations in the composite and thus assume a rigid-plastic response. The incompressibility assumed for the plastic part of the deformation and compatibility implies that the deformation in both phases is identical to the macroscopically applied deformation. Based on the deformation gradient tensor in \eqref{eq:model:def} (characterized by the stretch factor $\lambda$) the following plastic deformation rate tensor is obtained:
\begin{equation}
  \bm{D}_\mathrm{p}^k = \frac{\dot{\lambda}}{\lambda} \big( \vec{e}_x \vec{e}_x - \vec{e}_y \vec{e}_y \big)
\end{equation}
where $k$ is used to denote each of the phases.

From the flow rule it is known that
\begin{equation}
  \bm{D}^k_\mathrm{p} = \tfrac{3}{2} \, \dot{\varepsilon}^k_\mathrm{p} \,   \frac{\big(\bm{\sigma}^d \big)^k}{\sigma^k_\mathrm{eq}}
\end{equation}
where $\big(\bm{\sigma}^d \big)^k$ is the deviatoric stress tensor in phase $k$ and the effective plastic strain rate
\begin{equation}
  \dot{\varepsilon}^k_\mathrm{p} = \frac{2}{\sqrt{3}} \frac{\dot{\lambda}}{\lambda}
\end{equation}
By substitution of the above and setting $\sigma^k_\mathrm{eq} = \sigma^k_\mathrm{y}$ we find
\begin{equation}
  \big( \bm{\sigma}^d \big)^k =  \frac{\sigma^k_\mathrm{y}}{\sqrt{3}} \left( \vec{e}_x \vec{e}_x - \vec{e}_y \vec{e}_y \right)
\end{equation}
The total stress tensor in phase $k$ then reads
\begin{equation}
  \label{eq:crit:stress}
  \bm{\sigma}^k =
  \sigma^k_\mathrm{m} \bm{I} + \big( \bm{\sigma}^d \big)^k =
  \sigma^k_\mathrm{m} \bm{I} + \frac{\sigma^k_\mathrm{y}}{\sqrt{3}}
  \left( \vec{e}_x \vec{e}_x - \vec{e}_y \vec{e}_y \right)
\end{equation}
with $\bm{I}$ the second order unit tensor.

Because perfect bonding has been assumed, at the interface the normal stress is continuous, i.e.\
\begin{equation}
  \vec{n} \cdot \bm{\sigma}^\mathrm{soft} \cdot \vec{n} = \vec{n} \cdot \bm{\sigma}^\mathrm{hard} \cdot \vec{n}
\end{equation}
Substitution of \eqref{eq:crit:stress} yields
\begin{equation} \label{eq:crit:lam}
  \sigma_\mathrm{m}^\mathrm{soft} + \frac{\sigma_\mathrm{y}^\mathrm{soft}}{\sqrt{3}} \cos 2\theta =   \sigma_\mathrm{m}^\mathrm{hard} + \frac{\sigma_\mathrm{y}^\mathrm{hard}}{\sqrt{3}} \cos 2\theta
\end{equation}
Furthermore the average, or macroscopic, hydrostatic stress
\begin{equation}
  \bar{\sigma}_\mathrm{m} = \big( 1 - \varphi^\mathrm{hard} \big)\, \sigma_\mathrm{m}^\mathrm{soft} +   \varphi^\mathrm{hard} \, \sigma_\mathrm{m}^\mathrm{hard}
\end{equation}
Using this equation to eliminate $\sigma_\mathrm{m}^\mathrm{hard}$, equation~\eqref{eq:crit:lam} results in an expression for $\sigma_\mathrm{m}^\mathrm{soft}$. Using $\bar{\sigma}_\mathrm{m} = 0$ (since the macroscopic deformation in \eqref{eq:model:def} is isochoric) this expression reads
\begin{equation} \label{eq:crit:hydros}
  \sigma_\mathrm{m}^\mathrm{soft} = \frac{\varphi^\mathrm{hard}}{\sqrt{3}} \big( \sigma_\mathrm{y}^\mathrm{hard} - \sigma_\mathrm{y}^\mathrm{soft} \big) \cos ( 2\theta )
\end{equation}
Since $\sigma_\mathrm{y}^\mathrm{soft} < \sigma_\mathrm{y}^\mathrm{hard}$ we can determine the sign of $\sigma_\mathrm{m}^\mathrm{soft}$ as a function of $\theta$; i.e.\ it is positive for $\theta = 0$ degrees and negative for $\theta = 90$ degrees.

Using the result above, we can now understand that indeed the orientation of the identified topological feature determines the sign of the hydrostatic stress at its center. To this end the feature is first regarded as a horizontal band with intermediate material properties to the soft and hard phase (i.e.\ $\sigma_\mathrm{y}^\mathrm{soft} < \sigma_\mathrm{y}^\mathrm{band} < \sigma_\mathrm{y}^\mathrm{hard}$), as is illustrated in Figure~\ref{fig:crit}(c). From \eqref{eq:crit:hydros}, with $\theta = \pi/2$ we find that the hydrostatic stress in the band is tensile, while it is compressive in the soft phase around it. The band actually consists of hard particles and a soft region, separated by vertical phase boundaries (see Figure~\ref{fig:crit}(d)). This implies that the hydrostatic stress in the hard particles is lower than the band average and that the matrix element at the center experiences a higher (tensile) hydrostatic stress. A high tensile hydrostatic stress in the central soft element result, rendering it susceptible for damage. Note that if the applied deformation is reversed, or the geometry of the feature is rotated by $90$ degrees, the central soft element experiences compression and it is therefore unlikely to develop damage.

This simplified analysis highlights that in a realistic situation, such as an interrupted hard band, a small region of matrix phase only develops critical levels of damage when the band is orientated along the direction of applied positive strain.

\subsection{Influence of the neighborhood of the critical feature}

Among the numerous critical features present in all unit cells considered, the highest levels of damage occur by the interplay of different topological influences stemming from the area surrounding the feature, leading to the distribution shown in Figure~\ref{fig:ref:prob}. We illustrate the governing influences in this section by employing idealized deterministic microstructures. In these idealized microstructures, first a critical feature is placed in the (center of) the cell. We limit ourselves to the configuration in which a band of hard phase particles (spanning the entire width of the cell) is interrupted by one matrix element, as this appears to be the most critical case. Bands of soft matrix phase of specific orientation are then introduced, as discussed below. Finally, the remainder of the cell contains a third, uniform phase the properties of which are determined by a rule of mixtures, i.e.\ the yield stress of this third, mixture phase at a certain effective plastic strain equals
\begin{equation}
  \label{eq:crit:taylor}
  \sigma_\mathrm{y}^\mathrm{mixture} =
  \big( 1-\varphi^\mathrm{hard} \big)\, \sigma_\mathrm{y}^\mathrm{soft} +
  \varphi^\mathrm{hard}\, \sigma_\mathrm{y}^\mathrm{hard}
\end{equation}
where $\varphi^\mathrm{hard} = 0.25$; $\sigma_\mathrm{y}^\mathrm{soft}$ and $\sigma_\mathrm{y}^\mathrm{hard}$ are the yield stresses of the soft and hard phase. This mixture phase is used to model the surrounding `composite' of hard and soft phases. It is used merely to eliminate the effect of randomness in this part of the cell, while preserving a certain degree of contrast with the properties of the soft and hard phases. The use of the simple rule of mixtures results in some degree of approximation and as a consequence results cannot directly be compared to those of the fully randomized cells.

\begin{figure}[htp]
\centering
\includegraphics[height=130mm]{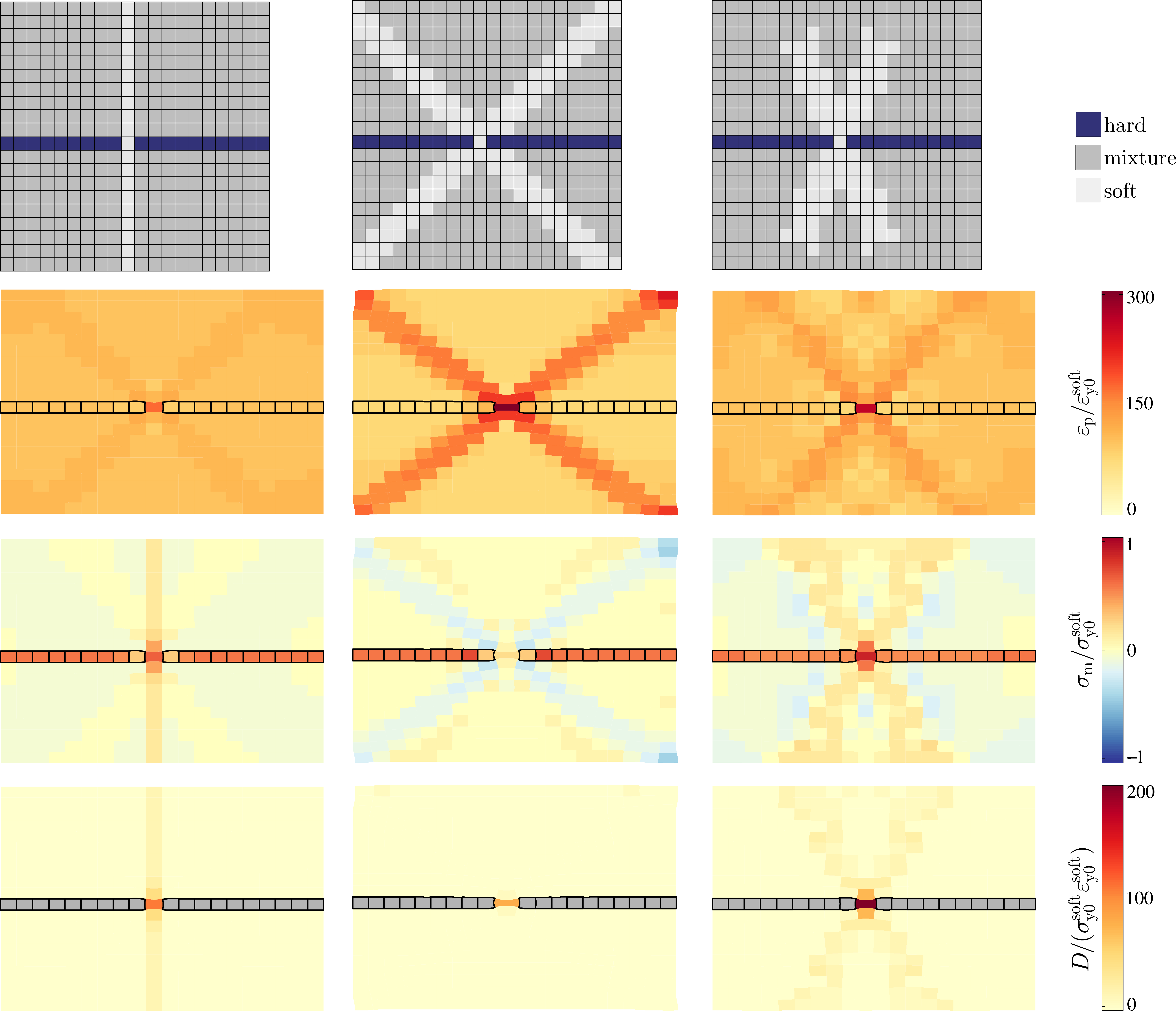}
\\ \vspace{0.3eM}
\footnotesize%
\begin{minipage}{38mm}
  (a)%
\end{minipage}%
\hspace{7.5mm}%
\begin{minipage}{38mm}
  (b)%
\end{minipage}%
\hspace{7.5mm}%
\begin{minipage}{38mm}
  (c)%
\end{minipage}%
\hspace*{20mm}%
\caption{Response of regularized microstructures, which display (a) stress dominated damage, (b) strain dominated damage, and (c) a critical combination of stress and strain. From top to bottom: the microstructure, normalized equivalent plastic strain $\varepsilon_\mathrm{p}$, hydrostatic stress $\sigma_\mathrm{m}$, and damage indicator $D$ at $\lambda = 1.2$.}
\label{fig:ref:ideal}
\end{figure}

In Figure~\ref{fig:ref:ideal}(a) the band of soft phase is oriented perpendicular to the interrupted band of hard inclusion phase. As explained in Section~\ref{sec:crit:mech} this configuration leads a high hydrostatic stress inside this band of soft phase. However, due to the absence of soft phase in the direction of deformation, the plastic strain (and therefore also the damage) remains small.

To introduce a high plastic strain, bands of the soft phase are oriented at favorable angles of $\pm 45$ degrees in Figure~\ref{fig:ref:ideal}(b). The response indeed shows a large amount of permanent deformation along these bands and in the central soft element. However, due to the orientation of the bands of matrix material, the stresses are compressive in the central soft element. For this reason, again, the damage indicator is low.

The most critical configuration is obtained by combining the effect of stress and strain. As observed in Figure~\ref{fig:ref:ideal}(c), this is accomplished by a trade-off between the configurations displaying high levels of stress and strain. The angle of the bands of soft phase is chosen under $\pm 55$ degree angles in the deformed configuration, thereby close to $\pm 45$ degrees to obtain maximum strain, while a positive hydrostatic stress is obtained. From the response of this configuration it is observed that due to the orientation of the bands of the soft phase, tensile stresses develop while permanent deformation is allowed. Quantitatively, the damage is larger than in any of the randomly generated microstructures. This configuration is similar to the distribution that appears in the probability of hard phase around the element displaying the highest damage, see Figure~\ref{fig:ref:prob}.

\section{Parameter study}
\label{sec:param}

Above the influence of the microstructural topology on damage has been investigated. In addition to the identified topological influence, there are other parameters which influence the damage. In this section, we consider the influence of two of these parameters: the volume fraction and hardness of the inclusion phase.

\subsection{Effect of the hardness}

To investigate the influence of the inclusion hardness, the right hand side in the hardening law \eqref{eq:model:yield} is multiplied by $\chi^\mathrm{hard}$ to obtain
\begin{equation}
  \sigma_\mathrm{y}^\mathrm{hard} =
  \chi^\mathrm{hard} \left( \sigma_\mathrm{y0}^\mathrm{hard} + H^\mathrm{hard}
  \varepsilon_\mathrm{p}^{n^\mathrm{hard}} \right)
\end{equation}
where $\sigma_\mathrm{y0}^\mathrm{hard}$, $H^\mathrm{hard}$ and $n^\mathrm{hard}$ are defined in \eqref{eq:model:param}. The factor $\chi^\mathrm{hard}$ is a scaling factor; for $\chi^\mathrm{hard} = 1$ the reference case considered above is recovered. We choose the microstructure identical to that in the cell in Figure~\ref{fig:ref:response}(c), displaying the highest damage at the final stage of deformation.

\begin{figure}[htp]
\centering
\includegraphics[height=95mm]{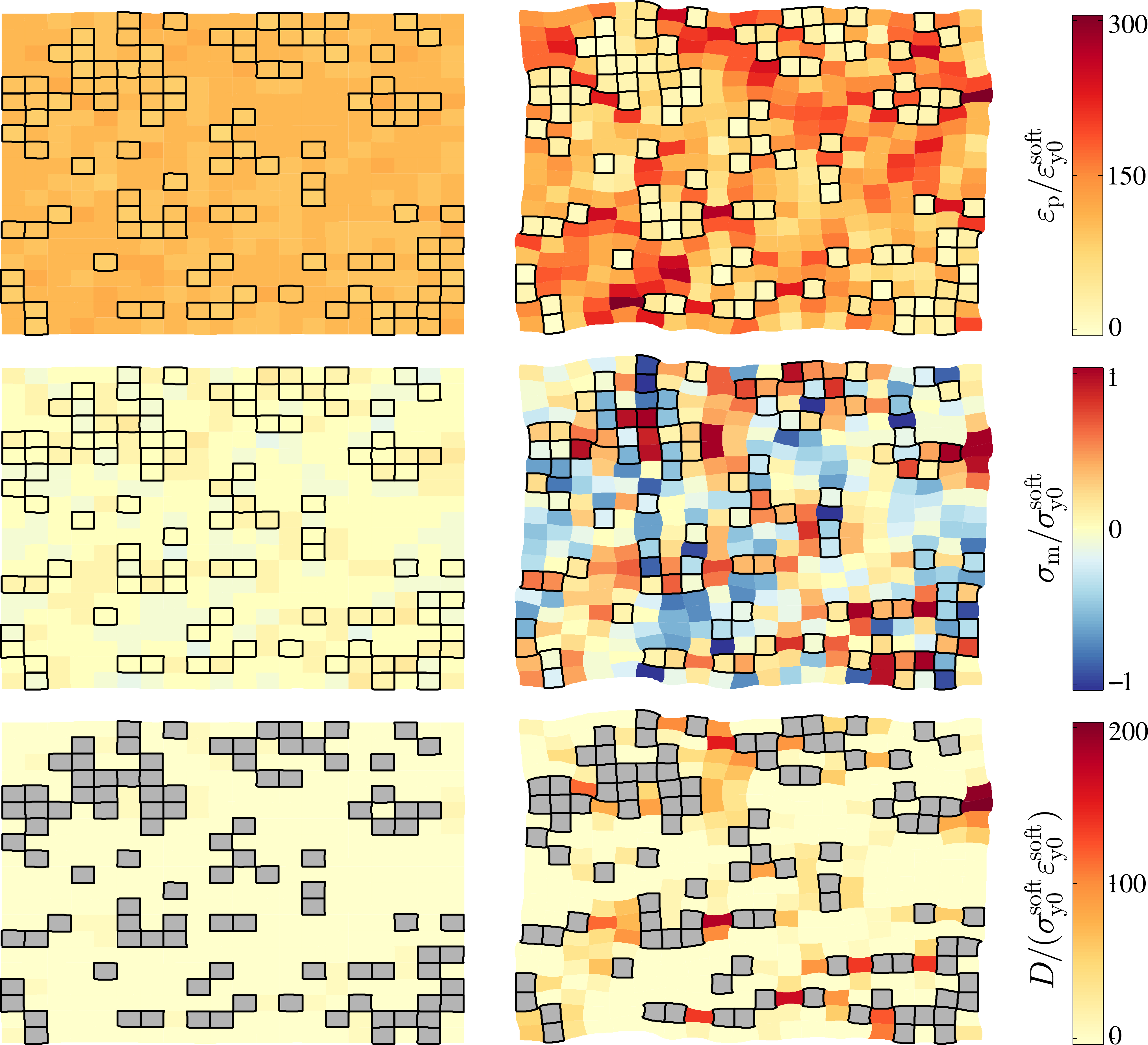}
\\ \vspace{0.3eM}
\footnotesize%
\begin{minipage}{38mm}
  (a) $\chi^\mathrm{hard} = 0.7$%
\end{minipage}%
\hspace{11mm}%
\begin{minipage}{38mm}
  (b) $\chi^\mathrm{hard} = 1.3$%
\end{minipage}%
\hspace*{20mm}%
\caption{Response for different hardnesses of the inclusion phase: (a) $\chi^\mathrm{hard} = 0.7$ and (b) $\chi^\mathrm{hard} = 1.3$, for the periodic cell of Figure~\ref{fig:ref:response}(c). From top to bottom: the normalized equivalent plastic strain $\varepsilon_\mathrm{p}$, hydrostatic stress $\sigma_\mathrm{m}$, and damage $D$.}
\label{fig:hard:response}
\end{figure}

The computed responses for two additional values of the inclusion hardness are shown in Figure~\ref{fig:hard:response}. For $\chi^\mathrm{hard} = 0.7$, in Figure~\ref{fig:hard:response}(a), the hardness of the inclusion phase is close to that of the matrix and the response is practically homogeneous. Compared to the reference, in Figure~\ref{fig:ref:response}(c), the damage indicator in the critical cell is reduced by $90\%$.

In Figure~\ref{fig:hard:response}(b), which corresponds to $\chi^\mathrm{hard} = 1.3$, the contrast between the properties of the phases is increased with respect to the reference case. Consequently the contrast in the stress and strain distributions increases as well and a much higher level of damage is reached at the same deformation (cf.\ Figure~\ref{fig:ref:response}(c)). In this case, the damage indicator in the critical cell in increased by $105\%$ compared to the reference in Figure~\ref{fig:ref:response}(c). Qualitatively the distribution of plastic strain, hydrostatic stress, and damage, are more or less unaffected by the change in $\chi^\mathrm{hard}$. The highest damage indicators are located in the critical feature identified above.

\begin{figure}[htp]
\centering
\begin{minipage}[b]{75mm}
  \includegraphics[width=1.\textwidth]{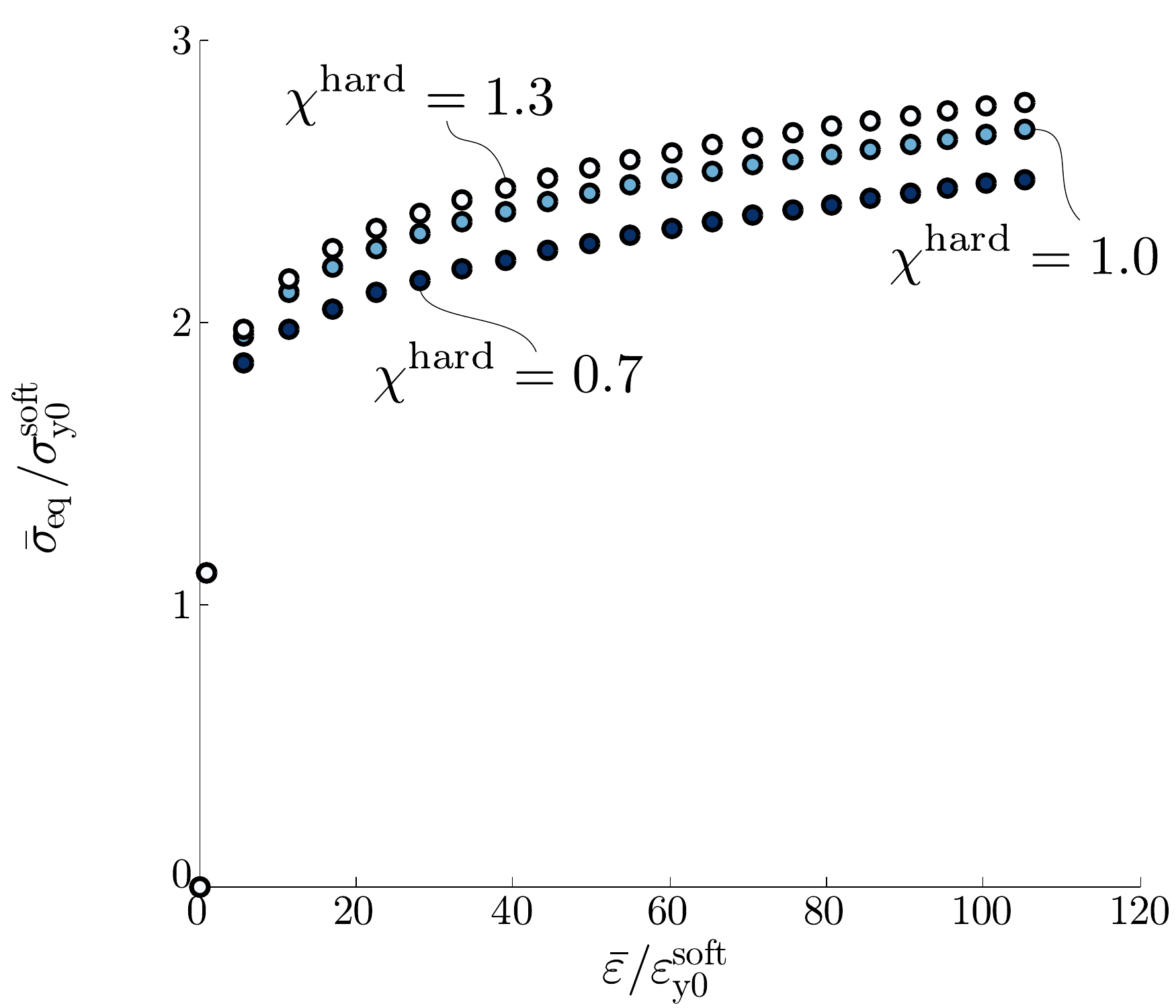}
  \\
  \hspace*{12.9mm}%
  \footnotesize (a) varying hardness
\end{minipage}
\hspace{5mm}
\begin{minipage}[b]{75mm}
  \includegraphics[width=1.\textwidth]{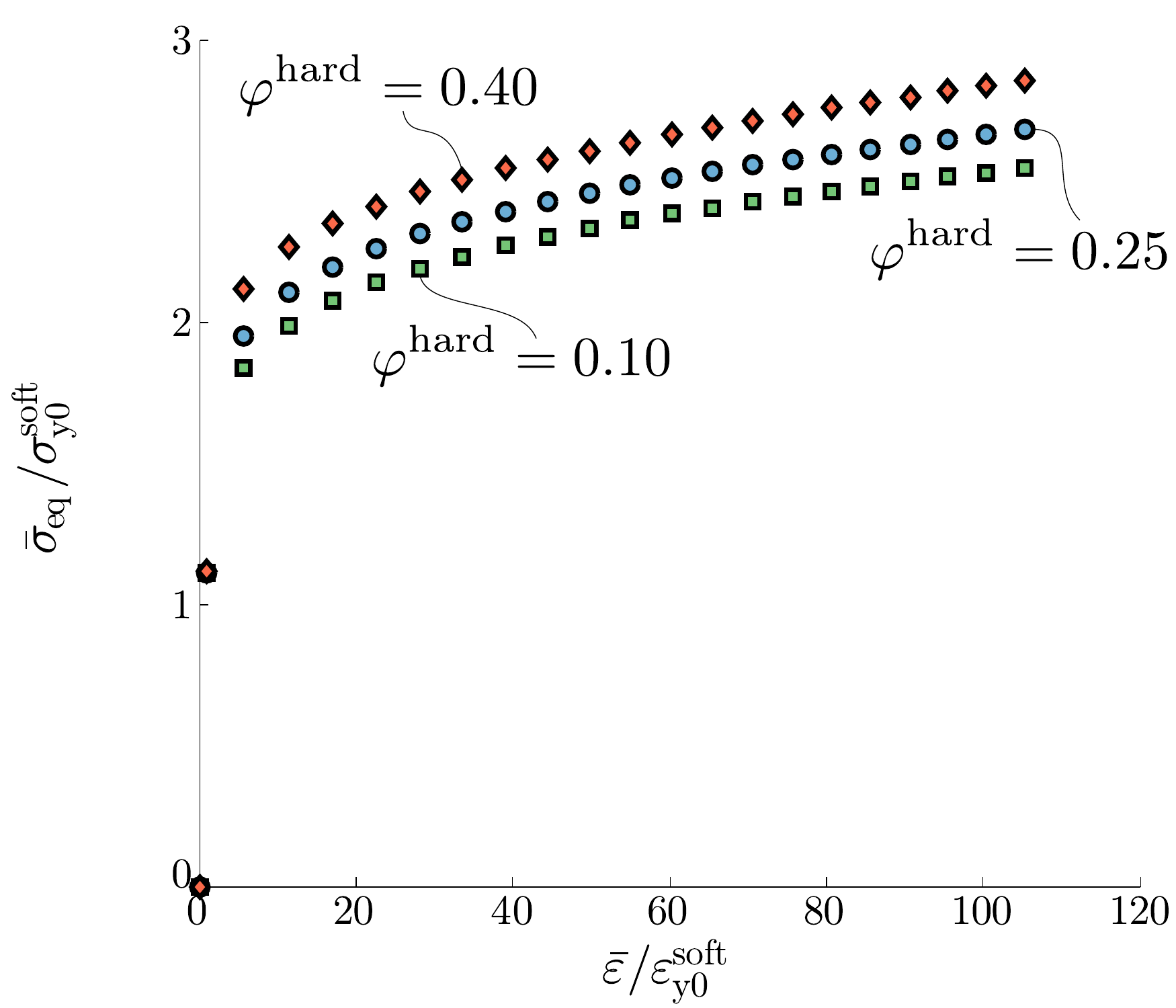}
  \\
  \hspace*{12.9mm}%
  \footnotesize (b) varying volume fraction
\end{minipage}
\caption{The macroscopic Von Mises equivalent stress $\bar{\sigma}_\mathrm{eq}$ as a function of the macroscopic applied equivalent (logarithmic) strain $\bar{\varepsilon}$ for varying (a) hardness of the inclusion phase characterized by the factor $\chi^\mathrm{hard}$, and (b) volume fraction of the inclusion phase $\varphi^\mathrm{hard}$.}
\label{fig:param:macros}
\end{figure}

When the macroscopic stress is studied it is found, as expected, that the macroscopic hardening increases with $\chi^\mathrm{hard}$, as is observed in Figure~\ref{fig:param:macros}(a). In this figure, the macroscopic Von Mises equivalent stress $\bar{\sigma}_\mathrm{eq}$ is shown as a function of the applied equivalent strain $\bar{\varepsilon}$ for the three different values of $\chi^\mathrm{hard}$ considered above ($0.7$, $1.0$, and $1.3$). Compared to the reference, for which $\chi^\mathrm{hard} = 1$, $\bar{\sigma}_\mathrm{eq}$ decreases $6\%$ for $\chi^\mathrm{hard} = 0.7$ and increases $4\%$ for $\chi^\mathrm{hard} = 1.3$ respectively at the same applied deformation.

\subsection{Effect of the volume fraction}

Next, the volume fraction of the inclusion phase, $\varphi^\mathrm{hard}$, is varied with the original reference value of the inclusion hardness (i.e.\ $\chi^\mathrm{hard} = 1$). In addition to the reference volume fraction of $\varphi^\mathrm{hard} = 0.25$ we consider $\varphi^\mathrm{hard} = 0.10$ and $0.40$. These values span the range of realistic values for e.g.\ dual-phase steel. For each volume fraction 400 randomly distributed cells are evaluated.

Similar to the hardness, changing the volume fraction affects the macroscopic response. This is shown in Figure~\ref{fig:param:macros}(b), which is representative for each cell with a different $\varphi^\mathrm{hard}$. The initial yield stress increases with the volume fraction, while the hardening is more or less constant (i.e.\ the curves are parallel). Specifically, for $\varphi^\mathrm{hard} = 0.40$ it increases by $6\%$ compared to the reference (width $\varphi^\mathrm{hard} = 0.25$), while for $\varphi^\mathrm{hard} = 0.10$ it decreases by $5\%$ at the final increment of deformation.

To compare the individual responses and the properties of onset of fracture for the different $\varphi^\mathrm{hard}$ the onset of ductile fracture is defined as a critical damage indicator value, which is arbitrarily taken as $D_\mathrm{c} = 80 \, \varepsilon_0^\mathrm{soft} \sigma_\mathrm{y0}^\mathrm{soft}$. The average strain and stress at this level of damage are defined as the fracture onset strain $\bar{\varepsilon}^\mathrm{f}$ and fracture onset stress $\bar{\sigma}_\mathrm{eq}^\mathrm{f}$ respectively.

\begin{figure}[htp]
\centering
\begin{minipage}[b]{75mm}
  \includegraphics[width=1.\textwidth]{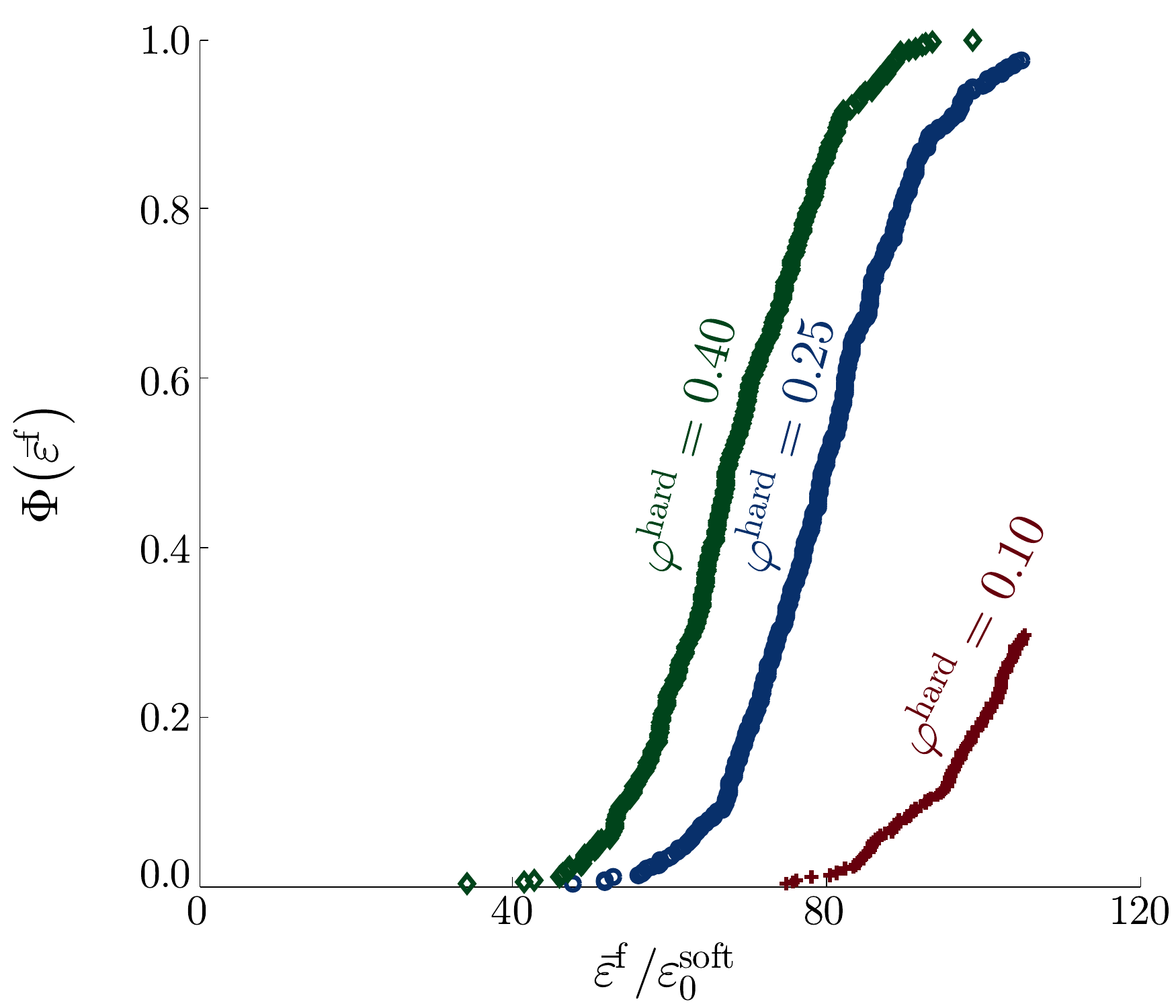}
  \\
  \hspace*{12.9mm}%
  \footnotesize (a) fracture onset strain
\end{minipage}
\hspace{5mm}
\begin{minipage}[b]{75mm}
  \includegraphics[width=1.\textwidth]{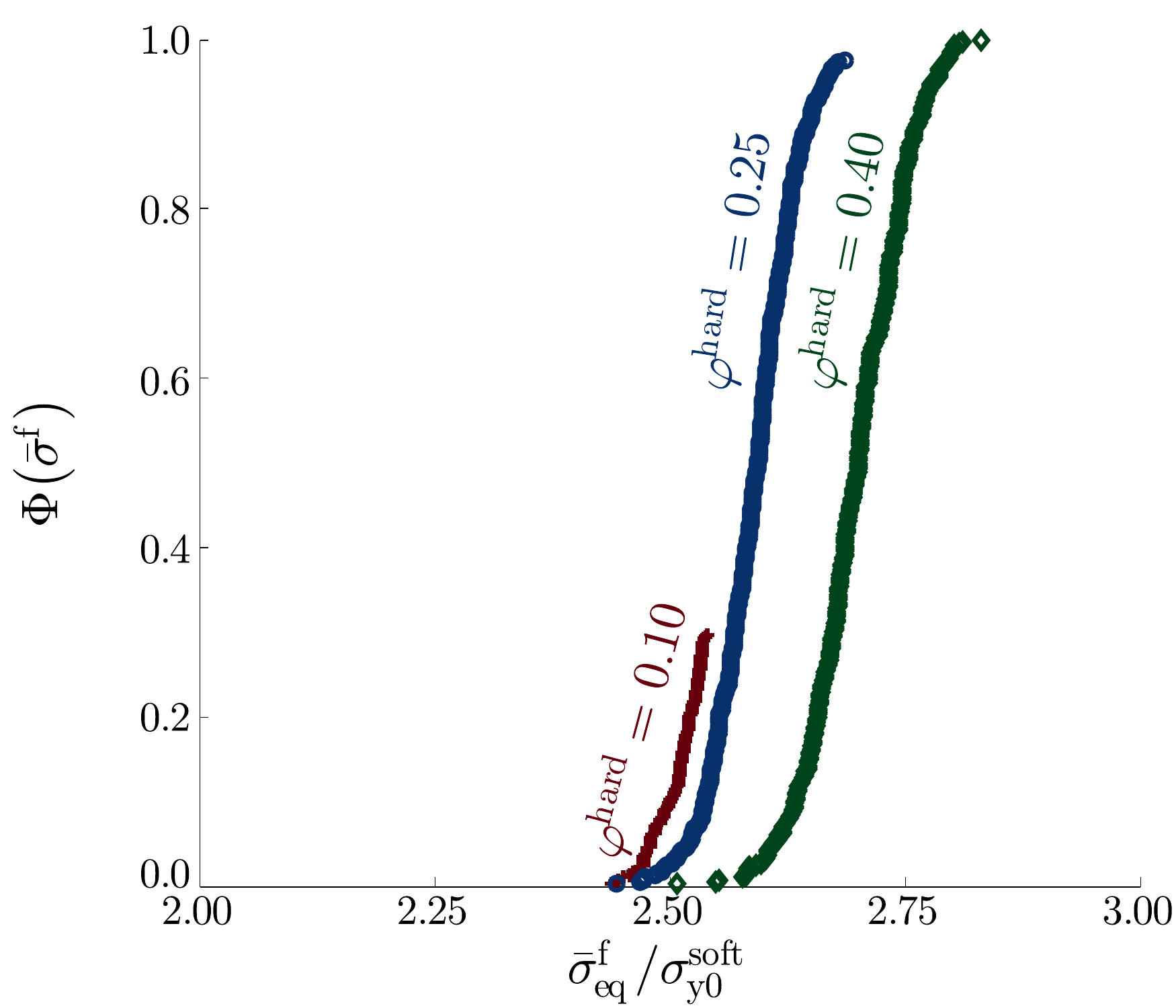}
  \\
  \hspace*{12.9mm}%
  \footnotesize (b) fracture onset stress
\end{minipage}
\caption{The cumulative probability $\Phi$ of the (a) fracture onset strain $\bar{\varepsilon}^\mathrm{f}$ and (b) fracture onset stress $\bar{\sigma}_\mathrm{eq}^\mathrm{f}$ as a function of their normalized values for 400 random realizations, for each volume fraction of hard inclusion phase $\varphi^\mathrm{hard} = 0.10$, $0.25$, and $0.40$.}
\label{fig:param:vol:cum}
\end{figure}

The cumulative distributions of the fracture onset strain and stress for all cells with different $\varphi^\mathrm{hard}$ are shown in Figure~\ref{fig:param:vol:cum}(a) and (b) respectively. For $\varphi^\mathrm{hard} = 0.10$ the critical level of damage is not reached in all cells at the final level of applied deformation. However, the focus is here given on the lowest part of the distribution, which is fully characterized. From Figure~\ref{fig:param:vol:cum}(a) it is observed that the fracture onset strain $\bar{\varepsilon}^\mathrm{f}$ decreases with increasing volume fraction of hard phase, $\varphi^\mathrm{hard}$, and that it is significantly affected by the microstructure. From Figure~\ref{fig:param:vol:cum}(b) it is observed that the fracture onset stress $\bar{\sigma}_\mathrm{eq}^\mathrm{f}$ among each of the 400 realizations is affected much less by the microstructure (notice the limited range on the horizontal axis in this diagram): it slightly increases with $\varphi^\mathrm{hard}$. These effects have been recognized in the literature, e.g.\ the effect of the martensite volume fraction in dual-phase steels \citep{Kim2000, Ahmad2000, Choi2009}, or the volume fraction of SiC particles in an aluminum matrix \citep{Llorca1991}.

\begin{figure}[htp]
\centering
\includegraphics[height=130mm]{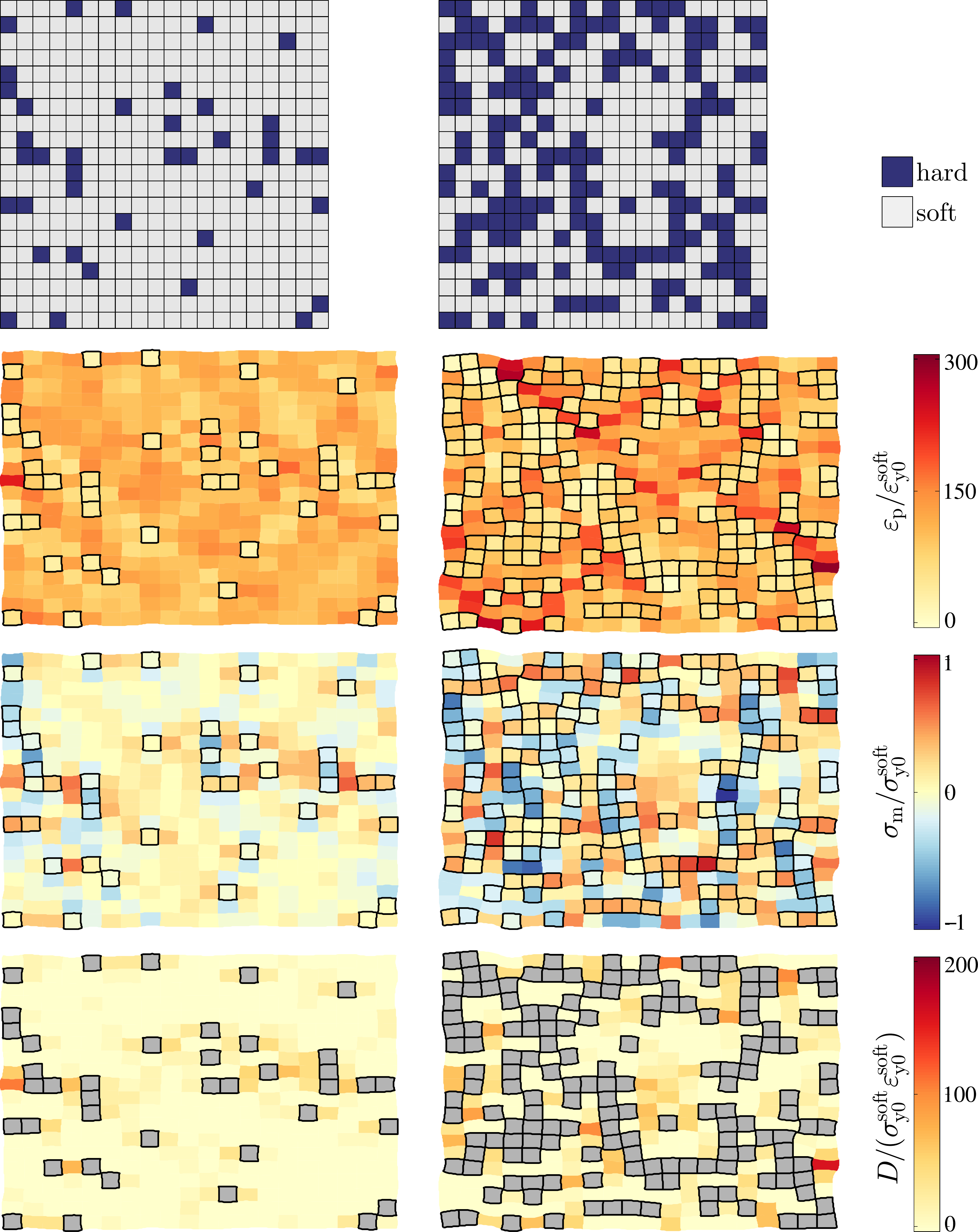}
\\ \vspace{0.3eM}
\footnotesize%
\begin{minipage}{38mm}
  (a) $\varphi^\mathrm{hard} = 0.10$%
\end{minipage}%
\hspace{9mm}%
\begin{minipage}{38mm}
  (b) $\varphi^\mathrm{hard} = 0.40$%
\end{minipage}%
\hspace*{18mm}%
\caption{Response of microstructures with different volume fractions of hard phase, $\varphi^\mathrm{hard}$ of (a) $0.10$, and (b) $0.40$. The cells shown are the ones for which the fracture onset strain $\bar{\varepsilon}^\mathrm{f}$ is lowest out of 400 realizations (for each $\varphi^\mathrm{hard}$). From top to bottom: the microstructure, the normalized equivalent plastic strain $\varepsilon_\mathrm{p}$, hydrostatic stress $\sigma_\mathrm{m}$, and damage indicator $D$.}
\label{fig:param:vol:response}
\end{figure}

The individual response is now analyzed for cells with the lowest fracture onset strain $\bar{\varepsilon}^\mathrm{f}$. For the reference case, where $\varphi^\mathrm{hard} = 0.25$, this response is shown in Figure~\ref{fig:ref:response}(c). For $\varphi^\mathrm{hard} = 0.10$ and $0.40$ the response is shown in Figure~\ref{fig:param:vol:response}(a) and (b) respectively.

For $\varphi^\mathrm{hard} = 0.10$, in Figure~\ref{fig:param:vol:response}(a), it is observed the plastic strain has a relatively low degree of heterogeneity. Only slightly elevated plastic strains appear inside the critical features (of Figure~\ref{fig:crit}(b)). A similar observation is made for the hydrostatic stress, whereby tensile stresses are found inside these features. Therefore also the damage indicator is elevated in these regions.

A different observation is made for $\varphi^\mathrm{hard} = 0.40$ (Figure~\ref{fig:param:vol:response}(b)). Here a heterogeneous distribution of the plastic strain is observed. A band of high strain is observed in the uninterrupted path of matrix at approximately $-45$ degrees (in the deformed configuration). Similarly, the hydrostatic stress is heterogeneous. Again, the damage indicator is highest inside the critical features.

Although not shown here, we have performed a similar statistical analysis as in Section~\ref{sec:ref:individual} for the different volume fractions. In the results a similar trend is observed. Consistently, the highest damage indicator is observed in the critical feature shown in Figure~\ref{fig:crit}(b). Furthermore, an increased probability of hard phase arises in the areas that are aligned with the primary deformation direction; bands of soft phase are observed under angles of 60 degrees.

\subsection{Combined effect}

Increasing either the hardness or the volume fraction of the hard phase has a similar effect, i.e.\ an increase in macroscopic stress but a reduction in ductility. This raises the question whether combinations of hardness and volume fraction exist which do not affect the macroscopic hardening response, but improve the ductility. To answer this question we have varied the hardness in a wide range ($0.7 \leq \chi^\mathrm{hard} \leq 1.3$) for different volume fractions ($\varphi^\mathrm{hard} = 0.325$, $0.40$) in addition to the reference case for which $\varphi^\mathrm{hard} = 0.25$ and $\chi^\mathrm{hard} = 1.0$. For each of these volume fractions, the unit cell for which the fracture onset strain $\bar{\varepsilon}^\mathrm{f}$ is lowest from all 400 randomly distributed cells has been selected.

\begin{figure}[htp]
\centering
\includegraphics[width=75mm]{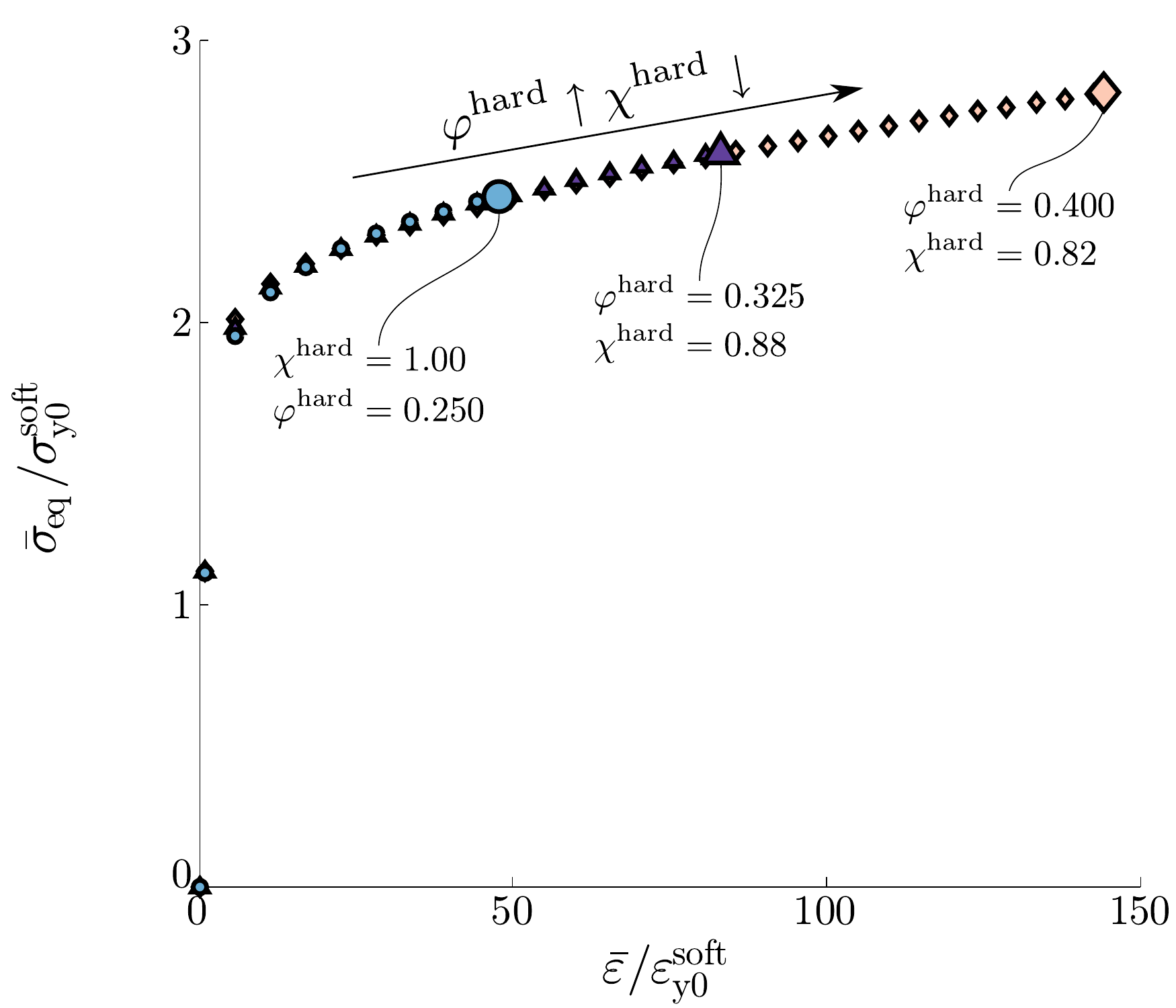}
\caption{Comparison of the onset of ductile fracture for different combinations of hard phase volume fraction $\varphi^\mathrm{hard}$ and hardness $\chi^\mathrm{hard}$, resulting in the same macroscopic hardening behavior. Different volume fractions are denoted by different markers. The onset of ductile fracture, i.e.\ point where each curve has been truncated, is highlighted using a bigger marker.}
\label{fig:param:macros:frac}
\end{figure}

In Figure~\ref{fig:param:macros:frac} the macroscopic stress--strain response for the reference case and the two higher hard phase volume fractions $\varphi^\mathrm{hard} = 0.325$ and $0.40$ are shown. The value of the hardness $\chi^\mathrm{hard}$ is chosen such that the macroscopic response is the same for all three combinations of $\varphi^\mathrm{hard}$ and $\chi^\mathrm{hard}$. Each of the curves has been truncated at the onset of ductile fracture, i.e.\ when the damage indicator reaches the critical value somewhere in the cell (see above). In the curves, this point has been highlighted. By increasing the volume fraction of the hard phase while at the same time decreasing the hardness, the fracture onset strain and fracture onset stress are increased significantly while the macroscopic hardening response remains unaffected. A similar conclusion can be found, although not explicitly, in the work of \citet{Choi2009} in which both the hardness and volume fraction of martensite are varied for dual-phase steel. \citet{Povirk1995} observes that when the contrast between the hard and soft phase is increased, the local plastic deformation in the soft phase increases. This is consistent with the present results, but neglects the influence of the (hydrostatic) stress on failure.

\section{Discussion}
\label{sec:discussion}

\subsection{Confrontation with experimental observations}

Based on the results obtained a critical topological feature in terms of the onset of ductile fracture has been identified, i.e.\ a small region of matrix phase flanked on both sides by hard phase particles which are aligned with the direction of applied positive strain (see Figure~\ref{fig:crit}).

\begin{figure}[htp]
\centering
\includegraphics[width=1.\textwidth]{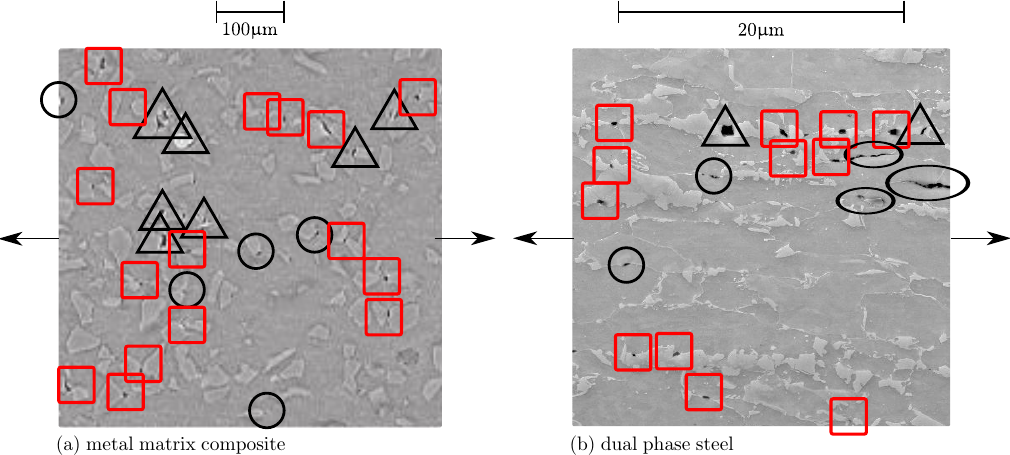}
\caption{Observations of damage incidents observed in (a) a SiC particle reinforced aluminum alloy matrix, and (b) a dual-phase steel. The pictures are cross-sections of (a) a specimen loaded in uniaxial tension, and analyzed using tomography (courtesy of \citet{Williams2010}); and (b) of a specimen loaded biaxially and analyzed using SEM (courtesy of Hoefnagels et al.\ \citep{Tasan2009, Hoefnagels2015}). In both pictures the direction of deformation has been indicated by arrows. The damage events similar to the identified critical feature (see Figure~\ref{fig:crit}) are highlighted by a red square, cracking of the hard particles or inclusion damage are indicated as black triangles, and other events by a black ellipse.}
\label{fig:exp}
\end{figure}

Experimentally, damage has been observed in such a feature on several occasions \cite[e.g.][]{Williams2010, Tasan2010, Hoefnagels2015}. \citet{Williams2010} performed a quantitative analysis of particle fracture and void growth in a metal matrix composite in which SiC particles are embedded in an aluminum matrix. An example of a reconstructed tomographic slice is shown in Figure~\ref{fig:exp}(a). In this figure, damage events similar to the critical feature of Figure~\ref{fig:crit} are highlighted by a red square, cracking of the hard particles or inclusion damage as black triangles, and other events by a black ellipse; the direction of deformation is indicated by arrows. Similarly, \citet{Hoefnagels2015} performed a quantitative analysis of damage in a dual-phase steel. Figure~\ref{fig:exp}(b) shows an image cross section made using scanning electron microscopy. From these two examples it is obvious that most of damage events in the soft phase occur in a feature similar to the critical feature in Figure~\ref{fig:crit}. Only some damage events in the soft phase occur in other regions. Cracking of the hard phase or inclusion induced damage is highlighted as another major damage mechanism, which is not included in the present analysis. Note that damage events different from Figure~\ref{fig:crit} might also be explained by the unknown sub-surface microstructure.

\subsection{Model approximations}

The simple microstructural model used in this study has provided valuable insights into the influence of the topology of the microstructure on damage indicator in the matrix phase of a ductile multi-phase material. In particular the highly idealized microstructure enabled a systematic analysis. For this purpose, a number of assumption had to be made. The influence of these assumptions is discussed next.

\begin{figure}[htp]
\centering
\includegraphics[width=.9\textwidth]{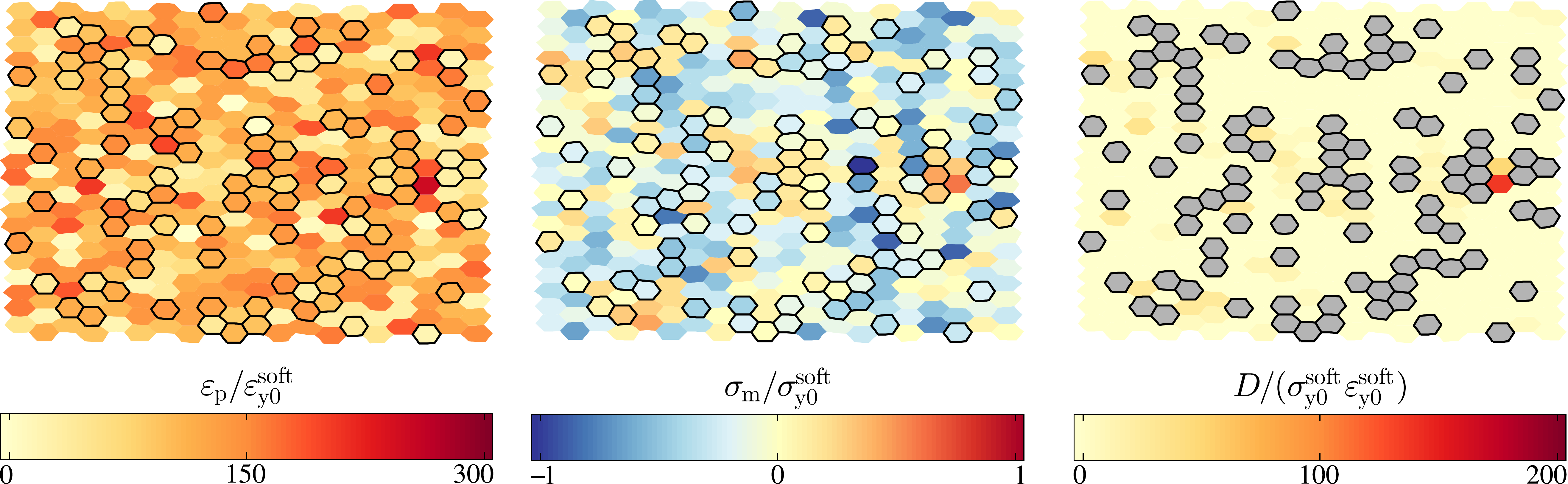}
\hspace{4mm}%
\caption{The response of a microstructure comprising hexagonal, instead of square, cells. From left to right: the normalized equivalent plastic strain $\varepsilon_\mathrm{p}$, hydrostatic stress $\sigma_\mathrm{m}$, and damage $D$; all taken at the stretch factor $\lambda = 1.2$.}
\label{fig:ref:hex}
\end{figure}

To generate the microstructure we have used square hard particles and square elements of matrix material. To focus on topological influences, we have considered only element averaged quantities. Doing so, the effect of geometrical singularities of the square shaped cells vanishes upon mesh-refinement. This is further confirmed by considering a different topology. To this end, a cell comprising hexagonal cells is considered. We again consider 400 randomly generated periodic cells with a volume fraction of $0.25$ of the hard phase, loaded in pure shear. Figure~\ref{fig:ref:hex} shows the response of the cell with the highest level of damage. In accordance with the above, we observe that damage is highest where hard inclusion phase particles are located left and right of a small region of matrix material, qualitatively similar to Figure~\ref{fig:crit}.

\begin{figure}[htp]
  \centering
  \includegraphics[width=.9\textwidth]{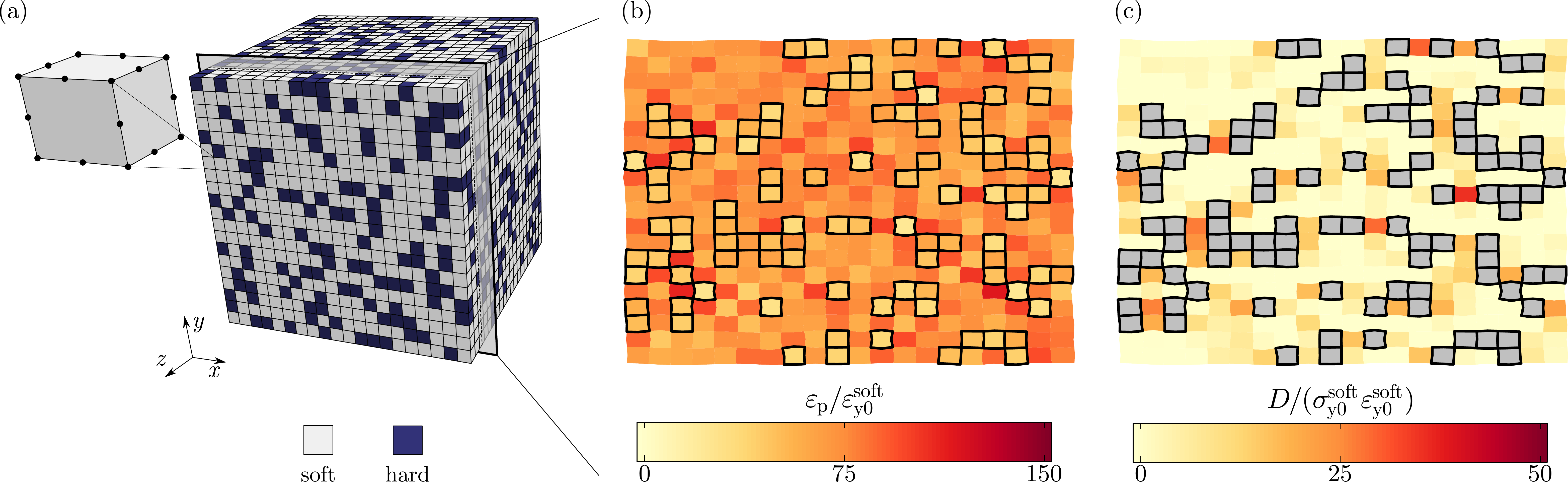}
  \hspace{4mm}%
  \caption{The mechanical response of a three-dimensional microstructure subject to pure shear (at the stretch factor $\lambda = 1.2$). From left to right: the microstructure, the normalized equivalent plastic strain $\varepsilon_\mathrm{p}$ and damage indicator $D$.}
  \label{fig:ref:plstress}
\end{figure}

A two-dimensional microstructure has been used throughout this paper. Both model and analysis can trivially be extended to three dimensions. However a systematic analysis that studies different three-dimensional deformation states using a sufficiently large unit cell and ensemble is outside the scope of this paper, and will be performed in a forthcoming contribution. Here, a preliminary analysis is performed to verify that the main conclusions are not dominated by restricting the model to two dimensions. Therefore, a full three-dimensional random microstructure, shown in Figure~\ref{fig:ref:plstress}(a), is subjected to the pure shear deformation according to \eqref{eq:model:def}. The same model, extended with periodicity conditions in all three spatial directions, is used to calculate the mechanical response. An additional approximation is made by discretizing each element using only one tri-quadratic finite element (see Figure~\ref{fig:ref:plstress}(a)). The mesh refinement study has shown that although this discretization leads to a solution that is qualitatively correct, quantitatively the response is overly stiff. The result is therefore only used only as a qualitative verification of the main conclusions. Of particular interest thereby is the arrangement of the phases around high damage. The presented result is therefore taken in a cross section parallel to the $xy$-plane in which damage is maximum. The response in terms of plastic strain and the damage indicator is shown in Figure~\ref{fig:ref:plstress}(b--c). In comparing the results to the earlier results in Figure~\ref{fig:ref:response}(c), it is observed the qualitative features are the same. In Figure~\ref{fig:ref:plstress}(b), the highest values of plastic strain are located in the soft phase in bands under $\pm 45$ degree angles. The clear difference is that the plastic strain in the hard phase is elevated compared with the two-dimensional case, which is an artifact of the chosen numerical discretization in the three-dimensional model. In Figure~\ref{fig:ref:plstress}(c), the highest values of the damage are located in the soft elements that have a hard element to the left and right and a soft element to the top and bottom, qualitatively similar to Figure~\ref{fig:crit}.

The influence of grain boundaries and grain orientation (in particular in the soft phase) has been disregarded. This made the identification of the distribution of the hard inclusion phase well defined. When the orientation would be taken into account, for instance using a crystal plasticity constitutive model, additional requirements on the orientation of the elements in the critical feature may be identified. Such an analysis only makes sense for a specific class of materials.

For the propagation of a crack (i.e.\ when the mechanical influence of damage is incorporated), the shape of the elements may be of influence. In this case, also the choice of a simple damage parameter indicating damage should be reviewed. Furthermore, the effect of element orientation may be more pronounced, and the assumption of an, isotropic, elasto--plastic constitutive model may be too restrictive as well.

\section{Conclusion}
\label{sec:conclusion}

This paper aimed to study the local morphology governing the onset of fracture of a ductile multi-phase material. The influence of the microstructural topology on the onset of ductile damage in the, relatively soft, matrix phase was analyzed. To this end we have used a highly idealized two-dimensional microstructural model in which topological influences are clearly identified. A large number of periodic cells have been investigated using a statistical framework making use of a simple damage indicator. We have identified the topological influences that lead to the highest damage indicator, i.e.\ the greatest probability of fracture. Besides the insights to fracture initiation, the presented framework has much potential to more systematically study the relation between microstructural morphology and fracture initiation. Indeed, the analysis may equally be applied in three dimensions, and one could also imagine calculating the average arrangement of phase around fracture initiation by directly using microscopic images.

This analysis has resulted in the identification of a single topological feature which is most sensitive to damage. In this feature, a small region of soft matrix material has hard inclusion phase particles on opposing sides; a similar observation was made by \citet{Kadkhodapour2011, Kadkhodapour2011a} for dual-phase steel. In the special case where additional hard phase particles are located on both sides, this feature resembles the interrupted bands frequently encountered in (industrially processed) dual-phase steel \citep{Tasan2010}.

The presence of the critical topological feature in the microstructure does not guarantee high damage. By analyzing the mechanics, it was found that such a configuration is only critical when aligned unfavorably with respect to the deformation. This is a consequence of the fact that the sign of the hydrostatic stress in the critical matrix element depends on the orientation of the phase boundaries. Specifically it was found that when the critical topological feature is rotated $90$ degrees (or the loading reversed) damage is always zero, due to the compressive stresses.

Besides the orientation, also the topology in the vicinity of the critical feature appears to have a significant influence on the indicated level of damage. Two mechanisms were identified. First, the number of hard phase particles along the direction of the maximum principal strain correlated with a more positive hydrostatic stress in the feature (i.e.\ more damage generally develops in a stronger hard phase band that is interrupted by a small region of matrix material). Second, a large amount of soft matrix in the direction of deformation causes high plastic deformation, and therefore damage. Also here, the orientation is essential. Depending on the orientation of a band of matrix material, either tensile or compressive stresses can develop respectively increasing or decreasing damage.

The analysis above was extended to address the influence of two other parameters: the volume fraction and hardness of the hard inclusion phase. It was found that the topological feature, discussed above, was critical regardless of these parameters. Analyzing the overall fracture onset strain and fracture onset stress using a critical damage indicator has shown that the fracture strain decreases with increasing volume fraction or hardness of the hard phase. At the same time, also the macroscopic hardening increases. By varying both parameters at the same time, for constant macroscopic stress--strain curve, an improvement of the predicted fracture onset strain (and, to a lower degree, the fracture onset stress) is obtained by increasing the volume fraction of the hard phase while at the same time decreasing its hardness. For dual-phase steel a similar observation has been made, using both experiments and models \citep{Sun2009a, Ahmad2000}.

It should be emphasized that in this paper, we have limited ourselves to modeling the onset of fracture of the ductile matrix and all of the above conclusions are based on this single failure mechanism. In reality, it is obvious that other failure mechanisms, such as e.g.\ fracture of the hard phase, contribute as well. We however believe that precisely by being able to `switch off' such additional mechanisms, which is not easily done in experiments, much additional insight can be obtained on a single mechanism -- whereby similar studies on the alternative mechanisms may be equally insightful.

\section*{Acknowledgments}

This research was carried out under project number M22.2.11424 in the framework of the Research Program of the Materials innovation institute M2i (\href{http://www.m2i.nl}{www.m2i.nl}).

J.P.M.~Hoefnagels (Eindhoven University of Technology) and F.~Maresca (Eindhoven University of Technology and M2i) are gratefully acknowledged for stimulating discussions.


\bibliography{library}

\end{document}